\documentclass[nohyper,11pt,letterpaper]{JHEP3}
\usepackage{graphicx}
\usepackage{amsmath}
\usepackage{amssymb}
\usepackage{amsthm}
\usepackage{amsfonts}

\newcommand{\be}{\begin{equation}}
\newcommand{\eb}{\end{equation}}

\newcommand{\ba}{\begin{eqnarray}}
\newcommand{\ab}{\end{eqnarray}}

\def\vev#1{\left\langle #1\right\rangle} 
\def\brn#1{\left( #1\right)} 
\def\abs#1{\vert #1 \vert}  

\newcommand{\vn}{{\bf \hat n}}
\newcommand{\vnn}{{\bf \hat n'}}
\newcommand{\vk}{{\bf \hat k}}

\newcommand{\stn}{\frac{\text{S}}{\text{N}}}

\title{Non-Gaussianities  in the Cosmological Perturbation Spectrum due to Primordial Anisotropy II}

\author{Anindya Dey\\
Theory Group, Department of Physics and Texas Cosmology Center\\ The University of Texas at Austin,
TX 78712.
\\ E-mail: \email{anindya@physics.utexas.edu}}

\author{Ely D. Kovetz \\
Theory Group, Department of Physics and Texas Cosmology Center\\ The University of Texas at Austin,
TX 78712.
\\ E-mail: \email{elykovetz@gmail.com}}
\author{Sonia Paban\\
Theory Group, Department of Physics and Texas Cosmology Center\\ The University of Texas at Austin,
TX 78712.
\\ E-mail: \email{paban@physics.utexas.edu}}

\abstract{ We continue to investigate possible signatures of a pre-inflationary anisotropic phase in two-point and three point correlation functions of the curvature perturbation for high-momentum modes which exit the horizon well after isotropization. The late time dynamics of these modes is characterized by a non-Bunch Davies vacuum state which encodes all the information about initial anisotropy in the background space-time. We observe that, unlike the non-planar momenta, there exist regimes of planar momenta for which scale invariance of the power spectrum is strongly broken. This regime of  planar momenta gives rise to enhanced non-Gaussianity in certain squeezed triangle configurations, although the enhancement of the $f_{NL}$ parameter is limited by the breakdown of linear perturbation theory at ``exact planarity". Finally, we demonstrate that for the range of planar modes for which scale invariance of the power spectrum is preserved, non-Gaussianity in the curvature perturbation spectrum is naturally constrained to be extremely small.}


\received{???????? ?st, 2011} \accepted{???????? ?th, 9999}
\preprint{\hepth{yymmnnn}\\UTTG-03-12 \\TCC-002-12 }

\begin{document}

\section{\bf Introduction}
In addition to the power spectrum, understanding detailed features of the inflationary scenario requires probing three-point correlation functions and beyond, collectively referred to as non-Gaussianitiy  \cite{Komatsu:2009kd}. It is well-known that observable non-Gaussianity  requires a departure from the standard single-field inflation with a canonical action \cite{Maldacena:2002vr, Creminelli:2004yq}.  Multiple scalar fields, non-canonical action for the scalar field, introducing higher derivative terms in the action or having a non-standard vacuum state (see  \cite{Bartolo:2004if} for reviews) are some of the variants from the standard inflationary model that have been studied.  A common feature of all these models, however, is that they have a homogeneous and isotropic background for the perturbations to evolve.\\
Investigating possible signatures of a strongly anisotropic phase of expansion at early times in the late time cosmological perturbation spectrum and its non-Gaussianity is an interesting problem in inflationary cosmology, given our total ignorance about the initial conditions. 
The first steps towards understanding the dynamics of such a scenario with a generic Bianchi I metric were undertaken in \cite{Pereira:2007yy, Gumrukcuoglu:2007bx} and the power spectrum was numerically determined.  In \cite{Kim:2011pt}, analytical expressions for the power spectrum of the scalar and tensor perturbations, in the case of an axially symmetric Bianchi I metric at early times, were derived for a regime of large momentum modes.\\ 
As discussed in \cite{Dey:2011mj, Kim:2011pt}, a particular axially symmetric Bianchi I background geometry (described in the next section) admits a WKB solution for perturbations at early times for modes in a certain high-momentum regime. On matching the WKB solution with the solution in the de Sitter phase, one can describe the late time dynamics of the curvature perturbation in terms of a non-standard ground state (essentially an excited state on the BD vacuum). Recent work \cite{Agullo:2010ws, Ganc:2011dy} claims that particular non-BD vacua can generate visible $f_{NL}^{local}$ for single-field inflation. The computation carried out in this work gives an example of a non-BD vacua for which this enhancement does not happen. 
Non-Gaussianity in the scalar perturbation spectrum for the non-planar modes in the high-momentum regime ($k_x \sim k_y\sim k_z \gg H_I$) were studied in \cite{Dey:2011mj}. \\
In this work, we primarily study non-Gaussianity in the scalar perturbation spectrum for another interesting high-momentum regime, viz. the planar modes where $ k_y \sim k_z \gg k_x$, where the $y$ and the $z$ axis define the plane of axial symmetry of the particular Bianchi I metric under study. We also analyze non-Gaussianities from three-point functions involving momenta in different regimes, viz. planar and non-planar. In addition, we study the power spectrum for modes in the planar regime and present an explicit analytical formula which explains several behaviors found numerically in \cite{Gumrukcuoglu:2007bx}\\

The paper is organized as follows. In the next section, we discuss the anisotropic background geometry in which the inflaton evolves and study the classical equations of motion. Section 3 deals with the study of cosmological perturbations in the high-momentum regime, the WKB solution and computation of the power spectrum in the two regimes mentioned. In section 4, we review the results for the bi-spectrum and the $f_{NL}$ parameter in the non-planar regime and in section 5 calculate the analogous quantities for the planar regime. We discuss the case of "mixed" modes (i.e. correlation functions involving planar as well as non-planar modes) in section 6 and present the conclusion in section 7.\\

\section{\bf The Model}
We consider a theory of Einstein gravity with a minimally coupled single scalar field given by the following action,
\begin{equation}
S=\frac{1}{2}\int d^4x \sqrt{-g} R + \int d^4x \sqrt{-g}\left(-\frac{1}{2}g^{\mu\nu}\partial_{\mu}\phi\partial_{\nu}\phi -V(\phi)\right), \hspace{3ex}(M_p^2\equiv 1)
\end{equation}
where the background metric is chosen to be an axially symmetric version of the
Bianchi I metric:
\begin{equation}
ds^2=-dt^2 + e^{2\rho}(dx)^2 +e^{2\beta}(dy^2+dz^2) \label{metric}
\end{equation}
In contrast to the FRW case where one has a single Hubble constant, we have two Hubble constants, which we choose to define as follows:
\begin{equation}
H=\frac{\dot{\rho}+2\dot{\beta}}{3}, \hspace{2ex} h=\frac{\dot{\rho}-\dot{\beta}}{\sqrt{3}}
\end{equation}
The classical dynamics of the system specified by the action (2.1) constitutes a strongly anisotropic expansion at early times (parametrized by $h$) followed by eventual isotropization at a time-scale $t \approx t_{iso}=\frac{1}{\sqrt{V}}$. For $t \gg t_{iso}$, the universe enters a phase of de Sitter expansion.\\

Note that $h$, which, roughly speaking, is a measure of the rate of anisotropic expansion vanishes in the isotropic limit ($\dot{\rho}=\dot{\beta}$) so that we are left with a single Hubble constant.\\

In terms of $H$ and $h$, the independent Einstein's equation and the equation of motion for the scalar field reduce to the following set of equations:
\begin{eqnarray}
\dot{H}+3H^2 &=&V(\phi)\\
3H^2-h^2 &=& \frac{1}{2}\dot{\phi}^2 + V(\phi)\\
\ddot{\phi}+3H\dot{\phi}+V'(\phi)& = & 0   
\end{eqnarray}

The time-evolution of $h$ can easily be derived from the above equations,
\begin{equation}
h(\dot{h}+3Hh)=0
\end{equation}
In the anisotropic phase, $h \neq 0$, which leads to the equation of motion,
\begin{equation}
\dot{h}+3Hh=0
\end{equation}

For a general $V(\phi)$, one can only obtain approximate solutions to the above system of equations. However, for a constant $V$, one can exactly solve the coupled differential equations for $H,h$ and $\dot{\phi}$ as follows,
\begin{eqnarray}
H& =& \sqrt{\frac{V}{3}} \, \coth({\sqrt{3V}t}) = H_I \,  \coth({\sqrt{3V}t}) \nonumber\\
h & = & \pm \sqrt{V} \, \frac{1}{\sinh({\sqrt{3V}t})}  \label{sip}\\
\dot{\phi} &= & 0  \nonumber
\end{eqnarray}

In the above solution, the constants have been chosen such that the metric approaches a Kasner solution in the limit $t \rightarrow 0^+$. The $\pm$ sign in the solution of $h$ indicates two different branches in the solution space (distinguished, among other things, by their behavior in the Kasner limit). It turns out that only for the positive branch, one can impose initial conditions on the cosmological perturbations at early times via the usual WKB approximation \cite{Gumrukcuoglu:2008gi}. Hence, in this note, we will focus exclusively on this class of backgrounds.\\

Now, for a given non-trivial $V(\phi)$, the slow-roll condition ($\ddot{\phi}
\approx 0$) will imply that at early times $V(\phi)$ is nearly constant with time, provided $H\dot{\phi}^2 \rightarrow 0$ at early times. This condition is obeyed by all common inflaton potentials and hence the above solution (\ref{sip}) can be trusted for a non-constant potential in the $t \rightarrow 0^+$ limit. As an example, consider $V(\phi)=\frac{m^2\phi^2}{2}$ for which $H$ and $\phi$ have the following asymptotic forms at early times,
\begin{eqnarray}
H & = & \frac{1}{3t}\left[1+\frac{m^2\phi_0^2t^2}{2}+O(m^4t^4)\right]\\
\phi & = & \phi_0 \left[1-\frac{m^2t^2}{4}+O(m^4t^4) \right] 
\end{eqnarray}
In this case, $H\dot{\phi}^2 \approx t \rightarrow 0$, so that $V$ is essentially constant at early times.\\

In this Kasner limit, the metric reduces to the following form,
\begin{equation}
ds^2_{Kasner}=-dt^2 + (\sqrt{V}t)^2 (dx)^2 + (dy^2 +dz^2)
\end{equation}
with $\dot{\rho}=\frac{1}{t}$, $\dot{\beta}=0$.\\

This is the gravitational background in which the cosmological perturbations
evolve at early times. The solutions for the background equations of motion suggest that the universe starts life with a very strong anisotropy ($h\rightarrow \frac{1}{t}$ at early times) which is smoothed out very fast by the inflaton potential. The universe then enters a phase of usual isotropic inflation.\\

\section{Power Spectrum: The Planar and the Non-planar regime}
The computation of the spectrum for cosmological perturbations for a generic anisotropic background has two significant differences with the corresponding computation in the isotropic case:\\
(1) The existence of a WKB solution for the modes of a given wavelength
at asymptotically early times ($t\rightarrow 0^+$) is not guaranteed , since 
for certain backgrounds any mode may become super-Hubble \cite{Pereira:2007yy} in this limit.\\
(2) The $SO(3)$ scalar and tensor perturbations are coupled for a generic wavelength at times $t \leq t_{iso}$ \cite{Gumrukcuoglu:2007bx}.\\

As commented in the previous subsection, the particular choice of the ``positive branch'' background solves (1). The positive branch metric in the Kasner limit is, in fact, a patch of Minkowski space-time \cite{Kim:2010wra} and this can be seen as follows:
Let $u=t \, \sinh({Vx})$ and $v=t \, \cosh({Vx})$, so that,
\begin{equation}
ds^2=-dt^2 + (\sqrt{V}t)^2 (dx)^2 + (dy^2 +dz^2)=-dv^2 + du^2 + (dy^2+dz^2) \label{metricMin}
\end{equation}
Therefore, this particular background could admit a WKB solution for perturbations at early times. Note that  coordinate invariants such as the curvature and the Weyl tensor are time independent and non vanishing and hence the space is never Minkowski. Thus, when computing quantities that depend on derivatives of the metric one should be careful if using (\ref{metricMin}).\\

 In \cite{Gumrukcuoglu:2007bx}, the perturbations in the anisotropic phase were parametrized in terms of the variables $V,H_{\times},H_{+}$ (which reduce to the usual gauge-invariant variables in the isotropic limit, $V$ becomes the Mukhanov variable and $H_{\times},H_{+}$ become the two polarizations of the tensor modes) with the following equations of motion:
\begin{eqnarray}
H_{\times}'' +\omega^2_{\times}H_{\times} & =& 0\\
\begin{pmatrix} V\\H_+ \end{pmatrix} ^{''} & =&- \begin{pmatrix} \omega^2_{11}& \omega^2_{12}\\\omega^2_{21}&\omega^2_{22} \end{pmatrix} \begin{pmatrix} V\\H_{+}  \end{pmatrix}
\end{eqnarray} 
where the derivative is with respect to the conformal time $\eta$ and the 
frequencies of the coupled system are given as,
\begin{eqnarray}
\omega_{11}^2 & = & e^{2\rho} \, (p_x^2+p_y^2- 2\dot{\rho}\dot{\beta}+\cdots)\\
\omega_{22}^2 & = & e^{2\rho}\, (p_x^2+p_y^2- 2\dot{\rho}\dot{\beta}+\cdots)\\
\omega_{12}^2 & = & e^{2\rho}\,  \left(\frac{\sqrt{2}p_y^2(\dot{\rho}-\dot{\beta})}{2\dot{\beta}p_x^2+(\dot{\rho}+\dot{\beta})p_y^2} \right) \, \left(-\frac{-3\dot{\beta}\dot{\phi}}{M_p}+\cdots \right)
\end{eqnarray}
The $p_i$ are the physical momenta, $p_x= k_1 e^{-2 \rho}, p_y= k_y e^{-2 \beta}$, where we have chosen the y-axis to coincide with the projection of the wavenumber/momentum vector on the plane of axial symmetry. The ellipsis in the above equations indicates terms subleading in the limit $t\rightarrow 0^+$. To the leading order at asymptotically early times, $\omega_{11}^2 \, e^{-2\rho},\omega_{22}^2\, e^{-2\rho} \approx 1/t^2$ while $\omega_{12}^2\, e^{-2\rho} \approx t^2$  and as a result, the mixing terms can be neglected. Therefore, in this limit, the scalar and tensor perturbations decouple as in the isotropic case. This feature is not surprising since, as seen earlier, the metric is that of a flat space-time.\\
The frequency of the decoupled $H_{\times}$ mode is given as.
\begin{equation}
\omega^2_{\times} = e^{2\rho} e^{4\beta}\left[ p_x^2+p_y^2 - \dot{\rho}(\dot{\rho}-\dot{\beta})+\dot{\beta} + {\dot{\phi}^2} + (\dot{\rho}-\dot{\beta})^2 \frac{p_x^2(p_x^2+4p_y^2) }{(p_x^2+p_y^2 )^2}\right] 
\end{equation}

\noindent Now, recall that in the isotropic case, the equation of motion of the Mukhanov variable $v$ is identical to that of  a scalar field $f$(after a redefinition $f \to a(\eta) f$) evolving in the same background, as long as the slow-roll conditions are obeyed (as a result of which $\frac{z''}{z} \approx \frac{a''(\eta)}{a(\eta)}$, with derivatives in terms of conformal time). In other words, the classical solutions of the scalar field $f$ and the curvature perturbation $\zeta$ are related as $f=\frac{\dot{\phi}}{H_I} \zeta$ in the slow-roll regime. Therefore, in all quantities of interest, one can substitute curvature perturbation by a solution of the  scalar field equation of motion, up to a well-defined normalization (specified above). In the anisotropic case, the scalar mode is given as 
$V=\exp{2\beta}[\delta\phi + \frac{p_y^2\dot{\phi}}{\dot{\rho}p_y^2+\dot{\beta}(2p_x^2+p_y^2)}\psi]$ in terms of the scalar field and metric fluctuation \cite{Gumrukcuoglu:2007bx}. Since $V$ approaches the usual Mukhanov variable in the isotropic limit, the above argument is also true for the scalar mode evolving in the anisotropic background in the limit $t \gg t_{iso}$ when the universe enters a late-time de Sitter phase.\\
At asymptotically early times, $V$  reduces to purely a fluctuation in the inflaton field in the Kasner background. In the limit $t \rightarrow 0^+ $,  $\frac{p_y^2\dot{\phi}}{\dot{\rho}p_y^2+\dot{\beta}(2p_x^2+p_y^2)} \approx t^2$, so that $V \approx e^{2\beta} \delta\phi$ (with $e^{2\beta} \approx constant$). If $f$ denotes a scalar field evolving in the Kasner background, then $V$ satisfies the same classical equation of motion as $e^{(\rho +2\beta)/3} f$ at early times as long as slow-roll conditions hold.\\
Thus, both at early and late times, the curvature perturbation can be understood in terms of this  scalar field $f$, evolving in the background given by the axially symmetric Bianchi I metric, which behaves as a Kasner metric at early time and de Sitter at very late times . Therefore, if one can find a WKB solution at early times, one can obtain an approximate classical solution  at late times by a standard matching procedure at some intermediate time. This approximate classical solution will specify a particular vacuum state of the curvature perturbation field at late times and can then be used to compute correlation functions in the ``in-in" formalism. It is important to note that, in this scheme, the entire information of anisotropy is encoded in the vacuum state of the theory.\\

We are interested in the solution of the above set of coupled equations for the perturbations at early times in the regime of high momentum : $|{\bf{k}}| \gg H_I$, which essentially implies that the modes are deep inside the horizon at early times and exit the horizon at times comparable to the time of isotropization or later (the case of planar modes is slightly more subtle, as explained later).\\

 As discussed in \cite{Kim:2011pt}, the high-momentum modes can be divided into two distinct categories depending on features of their early-time solutions:\\
\noindent (1) \textbf{Non-planar modes} i.e. $k_x \approx k_y \approx k_z  \gg H_I$ : A WKB approximation is valid at early times for this range of high-momentum modes and therefore the late time de Sitter solution for the perturbations can be obtained by matching this WKB solution at an intermediate time $t_{*}= \sqrt{\frac{k}{H_I}}$.\\
\noindent (2) \textbf{Planar modes} i.e. $k_y, k_z \gg k_x$ : The WKB condition is violated at very early times and restored at later times. In this case, one first finds a solution to the equation of motion in the regime where the WKB condition is violated and matches it to the WKB solution at a certain time $t_{\times}$ (see Appendix B). This WKB solution is then matched with the late time de Sitter solution at $t=t_*$ in a way similar to the non-planar case.\\ 

Note that in the non-planar case, the condition $|{\bf{k}}| \gg H_I$ implies that all the modes are sub-horizon at sufficiently early times. On the other hand, for planar modes with $k_x \sim 0$, the modes will be super-horizon in the x-direction no matter how early in time one goes back. For this regime of wave-numbers, one cannot impose initial conditions on the modes in a consistent way in our formalism. In fact, one can demonstrate that perturbation theory breaks down well before you reach this range of wave-numbers. In subsection 3.2, we will derive a rough lower bound for $k_x/H_I$ imposed by the validity of perturbation theory.\\

\subsection{Power Spectrum: Non-planar Mode}
In this subsection, we review the computation for the non-planar modes (for details, see \cite{Dey:2011mj}). As explained in Appendix A, the modes in this regime have a WKB solution at early times, which is used to obtain the precise solution in the de Sitter phase by matching at a time $t_{*}= \sqrt{\frac{k}{H_I}}$.\\
In the de Sitter phase, the classical solution to a scalar field $f$ is given as, 
\begin{equation}
f=\frac{A_{+} ({\bf k}) H_I}{\sqrt{2k^3}}\phi_{+}(\eta)+\frac{A_{-} ({\bf k}) H_I}{\sqrt{2k^3}}\phi_{-}(\eta) 
\end{equation}
where $\phi_{\pm}(\eta)=(k\eta \pm i)\exp{(\pm ik\eta)}$, with $\eta$ being the usual conformal time as defined in a de Sitter universe.\\
The coefficients $A_{+}$ and $ A_{-}$  are given as,
\begin{eqnarray}
A_{+} & =& \frac{1}{2}\left[(2-\varepsilon^2)+2i\varepsilon\left(\frac{\varepsilon^2}{2}-1\right) +O(\varepsilon^4)\right]\exp{\left(\frac{-i}{\varepsilon}\right)} \nonumber\\ \nonumber\\
A_{-} & = & -\frac{1}{2}\left[\left(\frac{2}{3}-r^2 \right)\varepsilon^3 +O(\varepsilon^4) \right]\exp{\left(\frac{i}{\varepsilon}\right)} \label{apam1}
\end{eqnarray}
where $\varepsilon=\sqrt{\frac{H_I}{k}}$ and $r=\frac{\sqrt{|k_y^2+k_z^2|}}{k}$.\\
The WKB approximation is valid when $\varepsilon \ll 1$  and we have retained terms up to order $\varepsilon^3$, which is the minimal order at which any signature of anisotropy appears.\\
>From the above solution, it immediately follows that,
\begin{equation}
|f|^2_{\eta \rightarrow 0} \longrightarrow \frac{H_I^2}{2k^3} |A_{+}-A_{-}|^2=\frac{H_I^2}{2k^3}\left[1+
\left(\frac{2}{3}-r^2 \right) \left(\frac{H_I}{k}\right)^{3/2} \cos\left(2\sqrt{\frac{k}{H_I}}\right)\right]
\end{equation}
Since the curvature perturbation is related to the scalar field $f$ as $f=\frac{\dot{\phi}}{H_I} \zeta$, the two-point correlation function of the curvature perturbation is given as,
\begin{eqnarray}
\left\langle \zeta_{\bf{k}}(t)\zeta_{\bf{k'}}(t)\right\rangle & \approx & (2\pi)^3\delta^3({\bf k}+{\bf k'}) \, \frac{{H_I}^2}{2k^3} \, \frac{H_I^2} {\dot{\phi}_e^2}\left[1+ \left(\frac{2}{3}-r^2\right)\left(\frac{H_I}{k}\right)^{3/2} \cos\left(2\sqrt{\frac{k}{H_I}}\right) \right] \nonumber \\ \\
& \equiv & (2\pi)^3\delta^3({\bf k}+{\bf k'}) \, \frac{F(k,\cos{\theta})}{2k^3}
\end{eqnarray}
where 
\begin{eqnarray}
\dot{\rho}(t_e) \, e^{\rho(t_e)} &\approx & k\\
\cos{\theta} &= & \frac{k_x}{k}= \sqrt{1-r^2} \\
\dot{\rho}(t_e) \approx H_I
\end{eqnarray}
The power spectrum obtained here is a slightly general form of \cite{Ackerman:2007nb} and admits the parametrization,
\begin{equation}
P(k)=P(k)_0 (1+\tilde{g}(k)+g(k)(\hat{k}.\vec{n})^2) \label{ACW}
\end{equation} 
where $P(k)_0$ is the usual nearly scale-invariant contribution while $\vec{n}$ is an unit vector in the direction which breaks the rotational invariance.\\

The spectral index for the curvature perturbation is then given as,
\begin{eqnarray}
n_s=k\frac{d}{dk}\log[F(k,\cos{\theta})] & \approx &   \frac{1}{\dot{\rho}_e} \frac{d}{d t_e}\log[F(k,\cos{\theta})] \nonumber \\ & \approx &  2(\eta -3\epsilon) 
 + \left(\frac{1}{3}-\cos^2{\theta}\right)e^{-\rho_e}\sin{(2 e^{\rho_e/2})} + O(e^{-3 \rho_e/2}) \nonumber \\
\end{eqnarray}
Note that, $e^{-\rho_e} \sim e^{-t_e/t_{iso}}$. One can easily estimate the magnitude of the correction term arising purely due to early-time anisotropy.\\
Here, the time $t_*$ (time at which we match the WKB solution with the de Sitter solution) obeys $t_* > t_{iso}$, such that $e^{\rho(t_*)} \gg 1$. In addition, it was shown in \cite{Dey:2011mj} (also explained in Appendix A) that $t_e \approx 2 t_*$. Therefore, for $t_* = 5 t_{iso}$, for example, we have $t_e = 10 t_{iso}$, which implies that the correction term is of the order $e^{-\rho_e} \sim e^{-t_e/t_{iso}} = e^{-10} \sim 10^{-5}$, while the slow-roll terms are of the order $10^{-2}$.
In this case, we are looking at a regime of momenta where $\frac{k}{H_e} \approx 10^5$ or, $k \approx 10^{-1} M_p$ ($H_e \sim 10^{-6} M_p$).\\
 The contribution of anisotropy to the two-point function could be larger for the regime of lower momenta (for which the WKB  approximation still holds). For $t_* = 2 t_{iso}$, we have $t_e = 4 t_{iso}$ which gives a correction term of the order of $e^{-t_e/t_{iso}} = e^{-4} \sim 1/50$ - comparable to the slow-roll terms. In this case,  $\frac{k}{H_e} \approx 50$, such that the WKB condition is still obeyed. However, since these modes have much larger  wavelengths, with $k \approx 5 \times 10^{-5} M_p$,  they may well be outside the present regime of observation. In any case, we will more be interested in the momentum regime where the contribution of anisotropy to the two-point function is negligible and investigate its possible observable signature in the three-point function.\\

\subsection{Power Spectrum: Planar Mode}
In this subsection, we study the two-point correlation function of the curvature perturbation and the corresponding power spectrum for the planar modes. As mentioned earlier and explained in Appendix B, imposing initial condition on these modes involves solving the classical equations of motion at early times when the WKB condition is violated. This solution is matched to the WKB solution at later times, which, in turn, is matched to the late time de Sitter solution.\\
As before, in the de Sitter phase, the classical solution to a scalar field $f$ is given as, 
\begin{equation}
f=\frac{A_{+} ({\bf k}) H_I}{\sqrt{2k^3}}\phi_{+}(\eta)+\frac{A_{-} ({\bf k}) H_I}{\sqrt{2k^3}}\phi_{-}(\eta) 
\end{equation}
where $\phi_{\pm}(\eta)=(k\eta \pm i)\exp{(\pm ik\eta)}$, with $\eta$ being the usual conformal time as defined in a de Sitter universe.\\
The coefficients $A_{+}$ and $ A_{-}$, obtained by matching procedure (explained in Appendix B), are given as,
\begin{eqnarray}
A_{+} & =& \frac{e^{i(\pi/4 - b k/H_I)}}{\sqrt{1-e^{-2\pi (a s k/H_I)}}} \nonumber\\ \nonumber\\
A_{-} & = &  \frac{e^{-\pi (a s k/H_I)}e^{-i(\pi/4 - b k/H_I)}}{\sqrt{1-e^{-2\pi (a s k/H_I)}}}\label{apam2}
\end{eqnarray}
where $s= k_x/k \ll 1$ is a measure of "planarity" of the mode while $a=\frac{2^{2/3}}{3}$ and $b=\frac{2^{2/3}\sqrt{\pi} \Gamma(1/3)}{3\Gamma(5/6)}$ are numerical constants. Note that $A_{\pm}$ obey the usual normalization condition, viz.,
\begin{equation}
|A_{+}|^2 - |A_{-}|^2 =1
\end{equation}
Given that the classical solutions of $f$ and $\zeta$ are related as, $f=\frac{\dot{\phi}}{H_I} \zeta$ , the two-point correlation function of the curvature perturbation is given as,
\begin{equation}
\left\langle \zeta_{\bf{k}}(t)\zeta_{\bf{k'}}(t)\right\rangle  \approx  (2\pi)^3\delta^3({\bf k}+{\bf k'}) \, \frac{{H_I}^2}{2k^3} \, \frac{H_I^2} {\dot{\phi}_e^2}\left[ \coth{(\pi a s k/H_I)}- \frac{\sin{(2bk/H_I)}}{\sinh{(\pi a s k/H_I)}}\right] 
\end{equation}
where 
\begin{eqnarray}
\dot{\rho}(t_e) \, e^{\rho(t_e)} &\approx & k\\
\dot{\rho}(t_e) \approx H_I
\end{eqnarray}

\noindent The power spectrum for the curvature perturbation then reads,
\begin{equation}
P(k)=P(k)_0\left(\coth{(\pi a s k/H_I)}- \frac{\sin{(2bk/H_I)}}{\sinh{(\pi a s k/H_I)}}\right)
\end{equation}
where $P(k)_0=\frac{{H_I}^2}{2k^3}\frac{H_I^2} {\dot{\phi}_e^2}$ is the power spectrum for standard inflation.\\
 It is useful to compare the above expression to the power spectrum obtained for the non-planar case. For a generic value of $s$, the power spectrum is not a perturbative modification of the standard power spectrum and differs non-trivially from the ACW formula (equation $\ref{ACW}$). In other words, the effect of anisotropy can be large even in the limit of high momenta in the planar regime and can strongly violate the scale-invariance of the spectrum, as we discuss below.\\
There are two interesting limits in which the power spectrum can be analyzed. Firstly, consider the limit $s \to 0$ (such that $\pi a s k/H_I \ll 1$ although $k/H_I \gg 1$). In this case, we have,
\begin{equation}
P(k)=P(k)_0 \left( \frac{1-\sin{(2bk/H_I)}}{\pi a s k/H_I} \right)=P(k)_0 \left(\frac{H_I}{\pi a k }\right) \left( \frac{1-\sin{(2bk/H_I)}}{s}  \right) \label{psplanar}
\end{equation}
It is instructive to compare the above equation to the behavior of the power spectrum found numerically in \cite{Gumrukcuoglu:2007bx} (equation (35)). For small enough $s$, the power spectrum precisely diverges as $1/s$, as predicted in \cite{Gumrukcuoglu:2007bx}. Moreover, for a given $k$, the range of $s$ for which this behavior is observed, shrinks to  smaller size as $k$ grows, simply because $s$ has to be progressively smaller in order to satisfy the condition $\pi a s k/H_I \ll 1$ as $k$ grows. This is also in agreement with the results obtained in \cite{Gumrukcuoglu:2007bx}. \\
The oscillatory behavior of the power spectrum (fig 3 of \cite{Gumrukcuoglu:2007bx}) for small $s$ (i.e. $s \ll H_I/k$) can also be analytically seen from the above formula. Smaller the value of $s$ larger are the oscillations of the power spectrum.  In the limit of large $k$, i.e. $s \gg H_I/k$,  the oscillations in the power spectrum die out and the spectral index approaches the scale invariant formula.\\

There is a strong bound on the planarity parameter $s$ due to the fact that perturbation theory has to be valid at early times and we derive this now. Since curvature perturbation can be understood in terms of the scalar field $f$ evolving in the Bianchi I background, perturbation theory will be valid if at all times the total energy of the scalar field is much smaller compared to the energy of the background, which (given that the metric only depends on time) implies,
\begin{equation}
\int d^3 x T^{00}_{(0)} \gg \int d^3 x T^{00}_{(f)}
\end{equation}
where $T^{00}_{(0)} $ is the background energy density and $T^{00}_{(f)} $ is the energy density due to the scalar field $f$. Note  that

\begin{eqnarray*}
\int d^3 x T^{00}_{(f)}&=&\int d^3 x \left(\frac{1}{2}\dot{f}^2 +\frac{1}{2}g^{ij}\partial_i f \partial_j f \right) \\
&=&\frac{1}{2}\int d^3 k \left[(\frac{d\tau}{dt})^2 f^{'}({\bf{k}},\tau) f^{'}(-{\bf{k}},\tau) + g^{ij} k_i k_j f({\bf{k}},\tau)f(-{\bf{k}},\tau)  \right] 
\end{eqnarray*}
where $\tau$ is the "conformal time" defined in Appendix B and prime denotes derivative with respect to  $\tau$. Now, plugging in the classical solution of $f$ at early times, i.e. $f({\bf{k}},\tau)= \frac{1}{\sqrt{6 a k_1}} \exp{[-3iak_1\tau +i\psi]}$ (also derived in Appendix B), we have,
\begin{equation}
\int d^3 x T^{00}_{(f)}=\frac{1}{2}\int d^3k \left[e^{-2\rho -4\beta} \frac{k_x}{6 a}(-9a^2 +e^{4\beta}) + \frac{e^{-2\beta}}{6 a k_x} (k^2_y +k^2_z) \right]e^{-6iak_x\tau + i\psi}
\end{equation}
where $\psi$ is a time-independent phase factor: $e^{i\psi}=e^{-iak_x/H_I \ln{(ak/H_I)}}$.\\
The only way to have a vanishing (or small) contribution from the above integral is to have a large exponent for the exponential factor so that integration over $k_x$ involves major cancellations. We therefore require, $|k_x \tau| \gg 1$. Since $\tau \approx \frac{1}{H_I}\ln{H_I t} $, this implies, $|\frac{sk}{H_I}\ln{H_I t}| \gg 1$. \\
Taking $H_I = 10^{-6} M_p$ and the minimum time-scale at which equations of General Relativity still holds as $t_{min}= 10 M^{-1}_p$, we obtain $\frac{sk}{H_I} \gg 1/10$. Therefore, the value of $s$ for which our formalism is valid must  obey the following ($k$-dependent) lower bound:
\begin{equation}
\frac{sk}{H_I} > O(1)
\end{equation}
The range of wave-numbers for which the power spectrum gives a $1/s$ behavior must satisfy $\frac{sk}{H_I} \ll 1$
and is therefore outside the physically sensible regime of wave-numbers.

The second interesting limit is $\pi a s k/H_I \gg 1$ (with $s \ll 1$), such that,
\begin{equation}
P(k) \approx P(k)_0
\end{equation}
where the isotropic behavior of the power spectrum is retrieved.\\
Finally, the spectral index for the curvature perturbation can be parametrized as $n_s= 2(\eta - 3\epsilon) + \Delta n_s$, where $\Delta n_s$ can be derived from the above equation as follows:
\begin{equation}
\Delta n_s= \frac{k}{H_I}(\frac{\pi a s \sinh{(\pi a s k/H_I)} - 2b \cos{(2bk/H_I)}}{\cosh{(\pi a s k/H_I)}- \sin{(2bk/H_I)}}-\pi a s \coth{(\pi a s k/H_I)})
\end{equation}
For $s > H_I/k$ such that $\sinh{(\pi a s k/H_I)} \sim \cosh{(\pi a s k/H_I)} \sim e^{\pi a s k/H_I}$, we have
\begin{equation}
\Delta n_s \approx -\frac{k}{H_I} e^{-\pi a s k/H_I} \cos{(2bk/H_I)} \label{deln}
\end{equation}
Equation $(\ref{deln})$ directly gives the regime of planar modes which preserve the scale invariance of the curvature perturbation spectrum,
\begin{equation}
e^{\pi a s k/H_I} \gg \frac{k}{H_I}
\end{equation}
 For a generic value of $s$ in this regime, the analytical formula for the spectral index suggests that the effect of the initial anisotropy might be substantial. However, in order to determine whether this effect will be observable, one needs to study the resulting statistical correlation between different multipoles of the CMB temperature anisotropies. In the next subsection, we numerically calculate the signal-to-noise ratio for individual multipoles for a physically reasonable range of values of $H_I$ as an estimate of how large  this effect can be. This analysis shows that the signal-to-noise ratio is pretty low even for the maximum allowed value for $H_I$. \\
 
 \subsection{Angular Correlation Matrix and Signal-To-Noise Ratio: Planar Modes}

Our starting point is the expression for the power spectrum obtained in the last subsection  
\ba\label{powspec}
\left<\zeta_k(t)\zeta_k(t')\right>&=&(2\pi)^3\delta^3({\bf k}+{\bf k}') P(k)_0\left(\coth(\pi ak \left|\cos{\theta}\right|)-\frac{\sin(2bk/H_I)}{\sinh(\pi ak \left|\cos{\theta}\right|)}\right)  \cr
&=&(2\pi)^3\delta^3({\bf k}+{\bf k}')P(k,\theta)
\ab
where $P(k)_0$ is the isotropic LCDM power spectrum and $\theta$ is the angle between $\bf{k}$ and the anisotropic direction of the Bianchi I metric.

We now want to connect this with the observable temperature anisotropies in the cosmic microwave background radiation.  
The full covariance matrix for the temperature anisotropies is given by 
\begin{equation}
  C_{\ell\ell' m m'} = \vev{a^{\,}_{\ell m}a^{*}_{\ell'm'}} = \int
  d\Omega d\Omega' \vev{\frac{\delta T}{T}(\vn,\eta_0,{\bf x}_0)\frac{\delta T}{T}(\vnn,\eta_0,{\bf x}_0)} Y_{\ell m}^{*}(\vn)Y^{\,}_{\ell' m'}(\vnn),
\end{equation}
where the temperature perturbation $\delta T/T(\vn,\eta_0,{\bf x}_0)= \sum_{\ell,m} a_{\ell m} \, Y_{\ell m} \brn{\vn}$ in the direction $\vn$ observed from ${\bf x}_0$ is related to the power spectrum of metric perturbations through
\ba
\frac{\delta T}{T}(\vn,\eta_0,{\bf x}_0) = \int \frac{d^3{\bf k}}{(2\pi)^3} \zeta ({\bf
  k},\eta_i)\Delta(k,\vk\cdot\vn,\eta_0)  e^{i{\bf k}\cdot{\bf x}_0} 
\label{dt2}
\ab
The transfer function $\Delta(k,\vk\cdot\vn,\eta_0)$ in fact depends only on the magnitude of the wavenumber $k = \abs{\bf k}$ and on the angle between ${\bf k}$ and the line of sight $\vn$, since the initial conditions for the evolution of perturbations after the onset of inflation can be considered isotropic \cite{GumConPel}. Using the line-of-sight approach \cite{SelZald} and omitting $\eta_0$ from now on for simplicity, we can expand in spherical harmonics 
\ba
\Delta(k,\vk\cdot\vn) &=& \sum_L (-i)^L (2L+1)
P_L(\vk\cdot\vn) \Delta_L(k) \nonumber\\
  P_L(\vk\cdot\vn) &=& \frac{4\pi}{(2L+1)} \sum_{M=-L}^L Y^{*}_{L
  M}(\vk)Y_{L M}(\vn)
\ab
and derive a generic expression for an anisotropic initial curvature power spectrum
\begin{eqnarray}
   C_{\ell\ell' m m'} &=& \int \frac{d^3{\bf k}}{(2\pi)^3}
 P({\bf k}) (4\pi)^2
 \sum_{LL'MM'}(-i)^{L-L'}\Delta_L(k)\Delta^{*}_{L'}(k)
 Y^{*}_{L M}(\vk)Y_{L' M'}(\vk)\nonumber\\ &&\times \int
 d\Omega d\Omega' Y_{\ell m}^{*}(\vn)Y^{\,}_{\ell'
 m'}(\vnn)Y_{LM}^{\,}(\vn)Y^{*}_{L' M'}(\vnn) \nonumber\\ 
  &=& (-i)^{\ell-\ell'}\frac{2}{\pi}\int d^3{\bf k}
P({\bf k})
 \Delta_\ell(k)\Delta^{*}_{\ell'}(k)
 Y^{*}_{\ell m}(\vk)Y_{\ell' m'}(\vk).
\end{eqnarray}

To simplify this expression for the axially symmetric power spectrum in our model, we use
polar coordinates in momentum space and rotate the system so that  $\theta$ is the angle
between ${\bf k}$ and $\hat{\bf z}$ (which now points in the anisotropic direction of the metric). Performing the $\phi$ integral and using $Y_{\ell m} (\theta,\phi) = \sqrt{\frac{(2\ell+1)(\ell-m)!}{4\pi(\ell+m)!}} P_\ell^m( \cos \theta )e^{im\phi}$ we get
\begin{eqnarray}
C_{\ell \ell' m m'} \!=\! 2\delta_{m m'} \left( - i \right)^{\ell - \ell'}\!\!\!  \int k^2 dk  \Delta_\ell ( k  ) \Delta_{\ell'}^{*}( k  )  \int\limits_{-1}^1 d(\cos\theta) Y_{\ell m} \left(  \theta,\phi\!=\!0 \right) Y_{\ell' m} \left(  \theta,\phi\!=\!0 \right) P( k, \theta) \;\;\;\;\;\;
\label{corrcllmm}
\end{eqnarray}

To calculate the signal-to-noise, we separate the isotropic and anisotropic contributions
\ba
C_{\ell m\ell'm'}\equiv   \vev{a^{\,}_{\ell m}a^\star_{\ell'm'}}=\delta_{\ell\ell'}\delta_{mm'}C_{\ell}^{(iso)}+\delta C_{\ell m\ell' m'}, \label{CMat}
\ab
where
$C_{\ell}^{(iso)}$ is the isotropic covariance matrix recovered by substituting $P( k , \theta) \equiv P \left( k \right)$
\ba
C_{\ell}^{(iso)} = \delta_{\ell \ell'} \delta_{m m'} C_{\ell \ell' m m'}^{({\rm iso})}=\frac{2}{\pi} \int k^2 dk P \left( k \right) \left|\Delta_\ell( k  )\right|^2
\ab
and $\delta C_{\ell m\ell' m'}$ is the deviation as a result of the anisotropic Bianchi I metric.
We start from the likelihood function for an isotropic LCDM covariance matrix 
 \cite{Heavens,Hamilton,Verde,Rathaus} for a system of $n$ Gaussian fields satisfying $\langle x_i x_j \rangle ={C}_{ij}^{(iso)}$
\ba\label{eq:likelihood}
{\cal L}^{(iso)}= \frac{1}{(2\pi)^{n/2}\sqrt{\det{{C}^{(iso)})}}} \exp{ \left(-\frac12 \mathbf{x}^{\text{T}} ({C}^{(iso)})^{-1} \mathbf{x}\right)}.
\ab
Given our perturbation ${C}=C^{(iso)}+\delta C$ we follow \cite{Rathaus,MasinaNotari,BenDavid} to define 
\ba
\chi^2(C)\equiv -2\left[\log({\cal L}(\text{C}))-\log({\cal L}^{(iso)})\right]
=\mathbf{x}^{{T}} \left( {C}^{-1} -({C}^{(iso)})^{-1}\right) \mathbf{x} +\log\left(\det {C} / \det {C}^{(iso)} \right).\;\;\;\;\;
\ab
The signal-to-noise for detecting the induced anisotropies is given by the mean value of $\chi^2$
\ba\label{eq:SN2}
\left(\stn\right)^2 \equiv \langle \chi^2(C)\rangle=\int \mathrm{d}x \chi^2(C)  {\cal L}^{(iso)}= \text{Tr} \left( {C}^{(iso)} {C}^{-1} -  1 \right) +\log\left(\det {C} / \det {C}^{(iso)} \right). \;\;\;\;\;
\ab

Expanding this to leading order in $\delta{C}$, we can examine the behavior for different multipoles $\ell$
\ba
\left(\stn\right)^{2}=\sum_{\ell}\left(\frac{\text{S}}{\text{N}}\right)^{2}_{\ell},
\ab
where (given the axial symmetry in our model)
\ba
\left(\stn\right)^{2}_{\ell}=\sum_{\ell'=2}^{\ell}\left(1-\frac{1}{2}\delta_{\ell\ell'}\right)\sum_{m=-\ell'}^{\ell'}\frac{|\delta C_{\ell m\ell'm}|^{2}}{C^{(iso)}_{\ell}C^{(iso)}_{\ell'}}.
\ab

\begin{figure}[h!]
\centering
\includegraphics[width=0.48\linewidth]{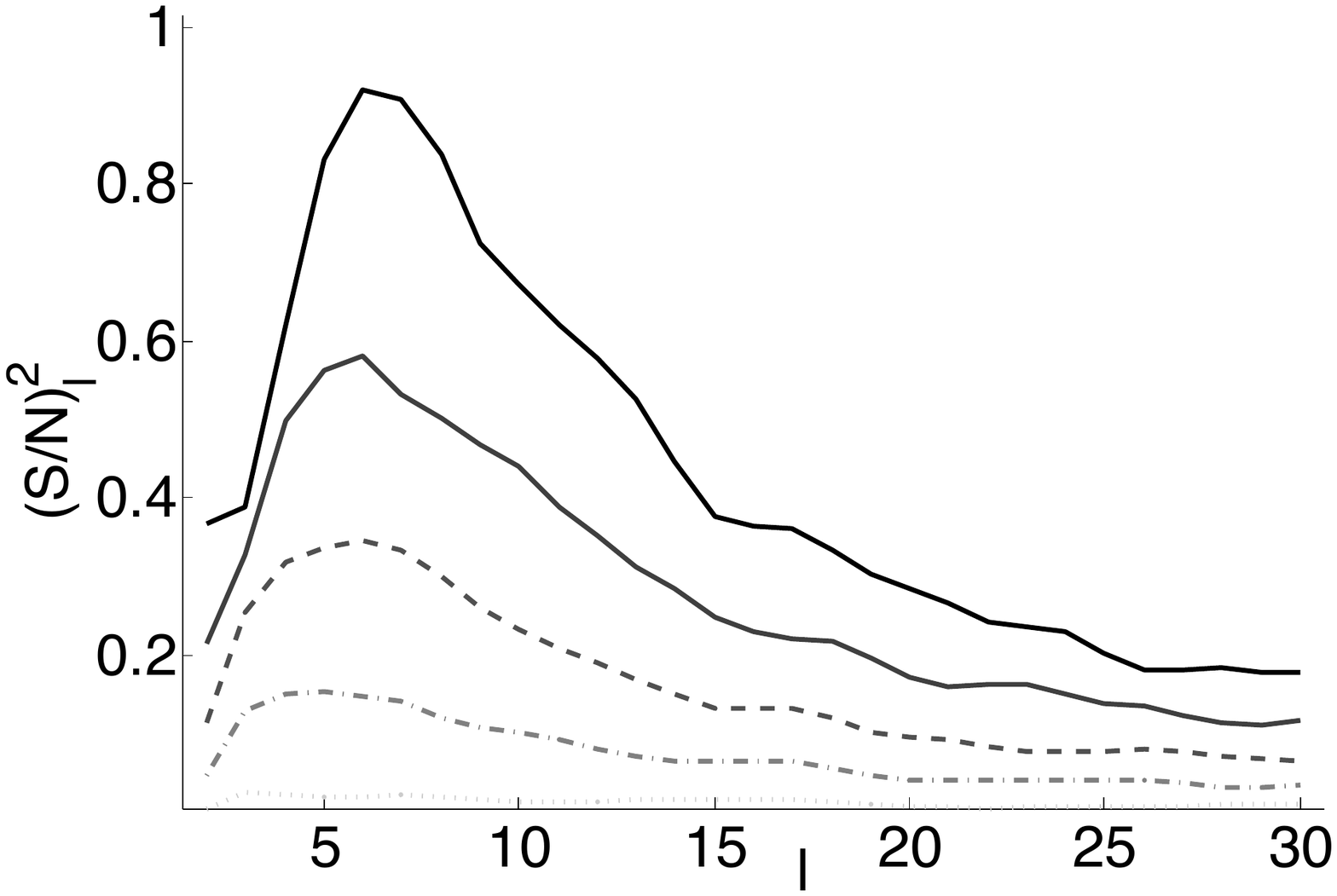}
\includegraphics[width=0.48\linewidth]{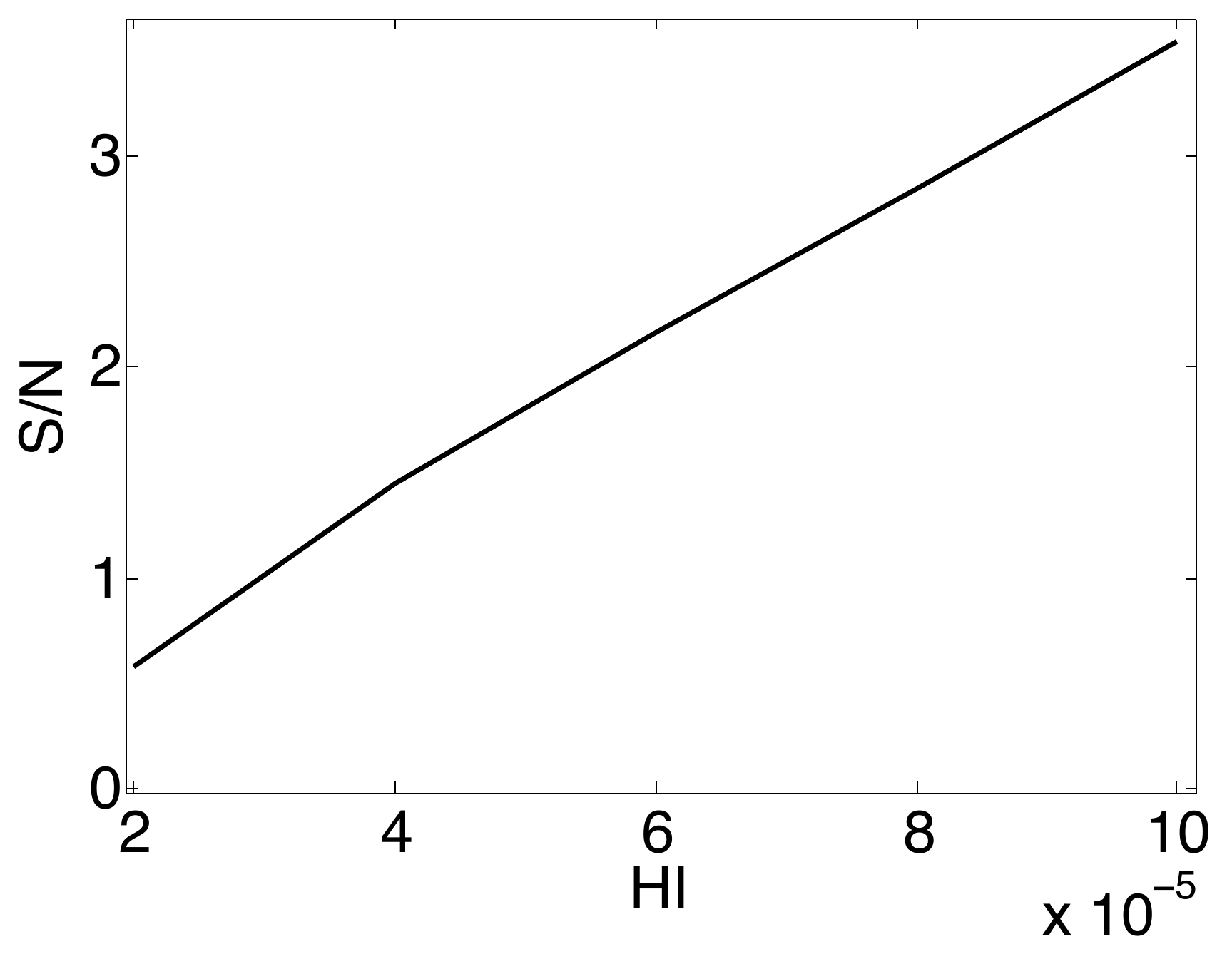}
\caption{{\it Left:} The ideal signal-to-noise per multipole $\ell$ for different values of $H_I$ between $10^{-5}M_{\rm pl}$ (yielding almost no signal) and $10^{-4}M_{\rm pl}$ (peaking at $l\sim7$ as a result of the tradeoff between cosmic variance and the signal dominance in large scales). {\it Right:} The total signal-to-noise in $\ell<30$ for different values of $H_I$. The upper bound $H_I\sim10^{-4}M_{\rm pl}$ yields a signal-to-noise ratio of $\sim3.5$ and will be challenging to detect in a real experiment.}
\label{SN_per_l}
\end{figure}

In Fig.~\ref{SN_per_l} we plot the signal-to-noise as a function of multipole $\ell$.\footnote{To avoid the breakdown of the perturbative treatment in the limit $\pi ak\cos{\theta}/H_I\ll1$, we cut off the power spectrum correction in (\ref{powspec}) at $\cos{\theta}=5H_I/(\pi ak)$ (verifying numerically that the results do not depend on this specific choice).}\footnote{We used transfer functions from CMBFAST \cite{SelZald} using CMBEASY \cite{CMBEASY} in the calculation.} The power spectrum correction affects small wave numbers $k$, yielding a signal in the largest scales, or smallest multipoles $\ell$, for which the statistics is more limited by cosmic variance.
We see that even for the maximum value allowed for $H_I$, the signal-to-noise is quite low and conclude that this feature in the power spectrum would most likely remain beyond experimental reach.



\section{Bispectrum: Non-planar modes}
In this section, we briefly review the results for the 3-point function of the curvature perturbation and the $f_{NL}$ parameter for  non-planar modes (for details, see \cite{Dey:2011mj} ).\\
For computing the 3-point function for the curvature perturbation, following  \cite{Maldacena:2002vr}, we consider a local (in time) non-linear field redefinition of $\zeta$:
\begin{equation}
\zeta=\zeta_c +\frac{\ddot{\phi}}{2\dot{\phi}\dot{\rho}}\zeta_c^2 + \frac{\dot{\phi}^2}{8\dot{\rho}^2}\zeta_c^2 + \frac{\dot{\phi}^2}{4\dot{\rho}^2}\partial^{-2}(\zeta_c\partial^2\zeta_c)
\end{equation}
Evidently, this redefinition does not change the quadratic action which implies that $\zeta_c$ and $\zeta$ have the same equation of motion and hence the same classical solution given by equation (3.1). Also, since local redefinitions do not yield any enhancement of the 3-point function, it is sufficient to compute the correlation function in terms of the redefined field $\zeta_c$. In terms of $\zeta_c$, the leading term (in slow-roll parameter) in the interaction Hamiltonian will be given as,
\begin{equation}
{\cal H}_I=-\int d^3x \, d\eta \, e^{3\rho} \left(\frac{\dot{\phi}}{\dot{\rho}}\right)^4  \dot{\rho} \,
\zeta_c^{'2}\partial^{-2}\zeta'_c \label{intHam}
\end{equation}
where the prime denotes derivative w.r.t. the conformal time $\eta$ (defined in the de Sitter phase), and the partial indicated space derivatives. \\
We can now use the ``in-in'' formalism to compute the tree-level contributions to the 3-point function, corresponding to the de Sitter vacuum specified by equation \ref{apam1}.\\
using the usual Feynman rules in the momentum space, the three-point correlation function at a conformal time $\eta$ is given as,
\begin{equation}
\left\langle \zeta_c({\bf k_1},\eta)\zeta_c({\bf k_2},\eta)\zeta_c({\bf k_3},\eta)\right\rangle_{R/L}\approx \delta^{(3)}({\bf k_1}+ {\bf k_2}+{\bf k_3})A^{R/L}({\bf k_1},{\bf k_2}, {\bf k_3},\eta)
\end{equation}
where $A^R$ and $A^L$ are given as,
\begin{eqnarray}
A^R({\bf k_1},{\bf k_2},{\bf k_3},\eta)=i\int ^{\eta}_{\eta_0}d\eta'e^{3\rho(\eta')}\left(\frac{\dot{\phi}}{\dot{\rho}}\right)^4\dot{\rho}\, \left(\sum^3_{i=1}\frac{1}{k_i^2}\right) \, \prod^{3}_{i=1}\partial_{\eta'}G_{{\bf k_i}}(\eta,\eta')\\
A^L({\bf k_1},{\bf k_2},{\bf k_3},\eta)=(A^R({\bf k_1},{\bf k_2},{\bf k_3},\eta))^{*}
\end{eqnarray}
where $G_{{\bf k_i}}(\eta,\eta')=\zeta_{cl}({\bf k_i},\eta)\zeta^{*}_{cl}({\bf k_i},\eta')$, with $\zeta_{cl}({\bf k_i},\eta)$ being the classical solution for curvature perturbation. $A^R({\bf k_1},{\bf k_2},{\bf k_3},\eta)$ denotes the contribution to the 3-point function corresponding to the tree-level Feynman diagram with a ``right'' vertex while $A^R({\bf k_1},{\bf k_2},{\bf k_3},\eta)$ denotes the contribution corresponding to the tree-level Feynman diagram with a ``left'' vertex. The final result for the 3-point correlation function of the curvature perturbation is given as,
\begin{equation}
\left\langle \zeta_c({\bf k_1},\eta)\zeta_c({\bf k_2},\eta)\zeta_c({\bf k_3},\eta)\right\rangle \approx \delta^{(3)}({\bf k_1}+ {\bf k_2}+{\bf k_3})[A^{R}({\bf k_1},{\bf k_2}, {\bf k_3},\eta) + A^{L}({\bf k_1},{\bf k_2}, {\bf k_3},\eta)]
\end{equation}
In the definition of $A^R({\bf k_1},{\bf k_2},{\bf k_3},\eta)$, one needs to make a choice of $\eta_0$ - which in standard inflationary scenario is taken to be $-\infty$. However, we will take $\eta_0$ to be of the order of the isotropization time-scale  when the universe has entered an essentially de Sitter phase, thus $\eta_0$ is near the onset of inflation and parallels the choice made in \cite{Holman:2007na} . In the computation of late-time correlation functions, the parameters $\dot{\rho}$ and $\dot{\phi}$  can therefore be assigned their respective de-Sitter values, which remain nearly constant during inflation.\\
In the limit of late times ($\eta \to 0$), we have,
\begin{equation}
A^R({\bf k_1},{\bf k_2},{\bf k_3})=-i\frac{\sum_{i<j}k^2_ik^2_j}{\dot{\phi}^2}
\int^0_{\eta_0}d\eta'\sum_{{\xi_i}=\pm1}\prod^3_{i=1}e^{i(\xi_ik_i)\eta'}  F_{\xi_i}(k_i)
\end{equation}
where the sum extends over all 8 possible linear combinations $\xi_ik_i$ and
$F_{\xi_i=-1}(k_i)=\frac{H_I^2}{2k^3}(|A^{i}_{+}|^2-A^{i}_{-}{A^{i}_{+}}^{\ast})$ and $F_{\xi_i=1}(k_i)=\frac{H_I^2}{2k^3}(|A^{i}_{-}|^{2}-A^{i}_{+}{A^{i}_{-}}^{\ast})$.\\
Therefore, on completing the $\eta'$ integration, we have,
\begin{equation}
A^R({\bf k_1},{\bf k_2},{\bf k_3})=-\frac{\sum_{i<j}k^2_ik^2_j}{\dot{\phi}^2}
\sum_{{\xi_i}=\pm1}(\prod^3_{i=1} F_{\xi_i}(k_i))\frac{1}{\sum_i\xi_ik_i}\left(1-e^{i \eta_0\sum_i\xi_ik_i}\right)  \label{aRf}
\end{equation}
where the functions $F_{\pm}(k_i)$ are given as,
\begin{eqnarray}
F_{\xi_i=1}(k_i)=\frac{\varepsilon_i^6}{4\dot{\rho}}\left(-\frac{1}{3}+\cos^2{\theta_i}\right)\varepsilon_i^3 \exp\left({\frac{-2i}{\varepsilon_i}}\right)\\
F_{\xi_i=-1}(k_i)=\frac{\varepsilon_i^6}{2\dot{\rho}}\left[1+\frac{1}{2}\left(-\frac{1}{3}+\cos^2{\theta_i}\right)\varepsilon_i^3 \exp\left({\frac{2i}{\varepsilon_i}}\right)\right]
\end{eqnarray}
There are two interesting configurations for which this amplitude is enhanced,\\
\noindent (1) \textbf{Flattened triangle limit} $k_2 \approx k_3 \approx k_1/2 \sim k $, such that $k_1=k_2 + k_3$. As discussed in \cite{Dey:2011mj}, the contribution of early time anisotropy to the $f_{NL}$ parameter is given as,
\begin{equation}
|\Delta f_{NL}|=\epsilon \, \left(\frac{\dot{\rho}}{k}\right)^{\frac{3}{2}}
\end{equation}
which, in addition to the standard slow-roll factor, is suppressed by powers of $ \frac{\dot{\rho}}{k}$, leading to an extremely small change in the value for $f_{NL}$.\\

\noindent (2) \textbf{Squeezed triangle limit}  $k_3\ll k_1 \approx k_2 \sim k$ but $|k_1-k_2|\neq k_3$. In this case, the change in the $f_{NL}$ parameter in the standard case is given as \cite{Dey:2011mj},
\begin{equation}
|\Delta f_{NL}| \approx \epsilon\left(\frac{\dot{\rho}}{k}\right)^{\frac{3}{2}}\frac{k}{k_3}
\end{equation}
which shows that the enhancement for the "squeezed triangle" limit is greater compared to the "flattened triangle" by a factor of $k/k_3$. However, since $(k/k_3)_{max} \sim 200$ given the range of scales measurable in the CMB, 
$|\Delta f_{NL}|$ is too small to be observed even for $k/H_I \sim O (50)$. For the range of wavelengths for which the contribution of anisotropy to the spectral index is negligible, the contribution is obviously far smaller.\\

\section{Bispectrum: Planar modes}
 In this section, we derive expressions for the $f_{NL}$ parameters for planar modes and study these in the regime $s > H_I/k$ , which includes the extreme limit
 $sk/H_I \gg 1$, where the scale-invariance of the power spectrum is retrieved.\\ 
The 3-point correlation function for the planar modes can be computed similarly in the ``in-in" formalism using the interaction Hamiltonian given in equation (\ref{intHam}) corresponding to the Bogoliubov transformed ground state specified by equation (\ref{apam2}). Following the same procedure  illustrated for the non-planar modes in the previous section, the ``right" amplitude can be written as,
\begin{equation}
A^R({\bf k_1},{\bf k_2},{\bf k_3})=-i\frac{\sum_{i<j}k^2_ik^2_j}{\dot{\phi}^2}
\int^0_{\eta_0}d\eta'\sum_{{\xi_i}=\pm1}\prod^3_{i=1}e^{i(\xi_ik_i)\eta'}  F_{\xi_i}(k_i)
\end{equation}
where the sum extends over all 8 possible linear combinations $\sum \xi_ik_i$ and
$F_{\xi_i=-1}(k_i)=\frac{H_I^2}{2k^3_i}(|A^{i}_{+}|^2-A^{i}_{-}{A^{i}_{+}}^{\ast})$ and $F_{\xi_i=1}(k_i)=\frac{H_I^2}{2k^3_i}(|A^{i}_{-}|^{2}-A^{i}_{+}{A^{i}_{-}}^{\ast})$.\\

Therefore, on completing the $\eta'$ integration, we have,
\begin{equation}
A^R({\bf k_1},{\bf k_2},{\bf k_3})=-\frac{\sum_{i<j}k^2_ik^2_j}{\dot{\phi}^2}
\sum_{{\xi_i}=\pm1}(\prod^3_{i=1} F_{\xi_i}(k_i))\frac{1}{\sum_i\xi_ik_i}\left(1-e^{i \eta_0\sum_i\xi_ik_i}\right)  \label{aRf2}
\end{equation}
Let us define $\tilde{F_{\xi_i}}(k_i)=\frac{2k^3_i}{H^2_I} F_{\xi_i}(k_i)$, which can be written as explicit functions of wavenumber using equation (\ref{apam2}) as follows:
\begin{eqnarray}
\tilde{F}_{-}(k_i)&=&\frac{(1- e^{-\pi (a s k_i/H_I)}e^{-i(\pi/2 - 2 b k_i/H_I)})}{1-e^{-2\pi (a s k_i/H_I)}}=\frac{e^{\pi a s k_i/H_I}+i e^{i2bk_i/H_I}}{2\sinh{(\pi a s k/H_I)} }\\ \nonumber \\
\tilde{F}_{+}(k_i)&=&-e^{-\pi a s k_i/H_I}e^{i(\pi/2 - 2 b k_i/H_I)} \tilde{F}_{-}(k_i)
\end{eqnarray}
As for the non-planar case, the amplitude computed above is enhanced for the flattened triangle configuration as well as the squeezed triangle configuration and are analyzed as follows:\\ \\

\noindent \textbf{Flattened Triangle Configuration}\\
 In this case, the enhancement appears when the wave-numbers satisfy $\sum \xi_{i} k_i=0$, so that the exponent of the exponential term in equation (\ref{aRf2}) vanishes.\\
We choose $k_2 \approx k_3 \approx k_1/2 $, setting $k_1=k_2 + k_3$. This implies that the set $\{\xi_i\}$ contributing to the enhanced bispectrum can have values $(1,-1,-1)$ and $(-1,1,1)$. Therefore
\begin{equation}
A^R({\bf k},{\bf k},{\bf k})=- \frac{3 H^6_I}{8 k^5 \dot{\phi}^2_e} \tilde{F}^3_{-}({\bf k}) \eta_0 (e^{-\pi a s k/H_I}e^{-i2 b k/H_I}- ie^{-2\pi a s k/H_I}e^{-i4 b k/H_I})
\end{equation}
Using the definition of $f^{flat}_{NL}$:

$$ f^{flat}_{NL}=(A^R({\bf k},{\bf k},{\bf k}) + c.c.)/P({\bf k})^2$$ we have
\begin{equation}
\Delta f^{flat}_{NL}=- \frac{\dot{\phi}^2_e}{H^2_I} \left(\frac{\cos{(2bk/H_I)}\cosh{(\pi a s k/H_I)} -\sin{(4bk/H_I)}}{(\cosh{(\pi a s k/H_I)}-\sin{(2bk/H_I)})^2}\right)
\end{equation}
In the $s > H_I/k$ limit,$$\Delta f^{flat}_{NL}\approx - \frac{\dot{\phi}^2_e}{H^2_I} e^{-(\pi a s k/H_I)}$$ which implies that the $f_{NL}$ parameter is suppressed by the exponential factor in addition to the slow-roll parameter. In the regime where the power spectrum is scale invariant, $e^{(\pi a s k/H_I)} \gg k/H_I $, we have $$|\Delta f^{flat}_{NL}| \ll \frac{\dot{\phi}^2_e}{H^2_I} \frac{H_I}{k}$$ This configuration, therefore, does not produce any appreciable enhancement in the $f_{NL}$ parameter in any regime.\\

\noindent \textbf{Squeezed Triangle Configuration}\\
For the squeezed triangle configuration, $k_3 \ll k_1 \approx k_2 \sim k$ and the corresponding amplitude is given as,
\begin{equation}
A^R({\bf k},{\bf k},{\bf k_3}) \approx -\frac{k^4}{\dot{\phi}^2}\frac{H^3_I}{8k^6 k^3_3} \sum_{{\xi_i}=\pm1}(\prod^3_{i=1} \tilde{F_{\xi_i}}(k_i))\frac{1}{\sum_i\xi_ik_i}\left(1-e^{i \eta_0\sum_i\xi_ik_i}\right) 
\end{equation}
where the sum is over the set   $\xi_i : (1,-1,-1),(-1,1,1),(1,-1,1),(-1,1,-1)$.\\
Let $s_3$ denote the planarity parameter for the smaller vector ${\bf k_3}$, while $s$ denotes the planarity parameter of the vector ${\bf k}$.
Using the definition of $f^{sqzd}_{NL}$: $$f^{sqzd}_{NL}=\frac{A^R({\bf k},{\bf k},{\bf k_3})}{P(k)P(k_3)} $$ we have 
\begin{eqnarray}
\Delta f^{sqzd}_{NL}=\left(\frac{\dot{\phi}^2_e}{H^2_I}\right) \left(\frac{k}{k_3}\right) \frac{1}{\sinh{\pi a s k/H_I} (\cosh{\pi a s k/H_I}-\sin{2bk/H_I})} \nonumber \\
\times \frac{1}{(\cosh{\pi a s_3 k_3/H_I}-\sin{2bk_3/H_I})}\left[ \alpha(k_3, \eta_0,s_3) + e^{\pi a s k/H_I} \, \beta(k,k_3,\eta_0,s_3)+ e^{-\pi a s k/H_I}\gamma(k,k_3,\eta_0,s_3)\right] \nonumber \\
\end{eqnarray}
where the functions $\alpha(k_3, \eta_0, s)$ , $\beta(k,k_3,\eta_0,s_3)$ and $\gamma(k,k_3,\eta_0,s_3)$ are given as,

\begin{equation}
\alpha(k_3, \eta_0, s_3)= -2(e^{\pi as_3 k_3}+ie^{i2bk_3/H_I})\left[(1-e^{i\eta_0 k_3}) +i e^{-\pi a s_3 k_3/H_I}e^{-i2bk_3/H_I}(1-e^{-i\eta_0 k_3})\right] +c.c.
\end{equation}
\begin{equation}
\beta(k,k_3,\eta_0,s_3)=i(e^{\pi as_3 k_3}+ie^{i2bk_3/H_I})\left[(1-e^{i\eta_0 k_3}) +i e^{-\pi a s_3 k_3/H_I}e^{-i2bk_3/H_I}(1-e^{-i\eta_0 k_3})\right] e^{-i2bk/H_I}+c.c.
\end{equation}
\begin{equation}
\gamma(k,k_3,\eta_0,s_3)=-i(e^{\pi as_3 k_3}+ie^{i2bk_3/H_I})\left[(1-e^{i\eta_0 k_3}) +i e^{-\pi a s_3 k_3/H_I}e^{-i2bk_3/H_I}(1-e^{-i\eta_0 k_3})\right] e^{i2bk/H_I}+c.c.
\end{equation}
As before, in the regime $s > H_I/k$, one can estimate the order of magnitude of the $f_{NL}$ parameter:
\begin{equation}
|\Delta f^{sqzd}_{NL}| \approx  \left(\frac{\dot{\phi}^2_e}{H^2_I}\right)\left(\frac{k}{k_3}\right) e^{-\pi a s k/H_I} \label{planarsqpure}
\end{equation}
Evidently, the squeezed triangle  configuration can lead to larger non-Gaussianity in this regime compared to the flattened triangle case. However, since $(\frac{k}{k_3})|_{max} \sim O(100)$, the enhanced $f_{NL}$ can at best be of $O(10^{-2})$. 

\section{Bispectrum: Mixed case}
We now consider the case where some of the momenta $\{\vec{k}_i\}$ belong to the non-planar regime while others are in the planar regime. As before, we will only compute the leading contribution to $f_{NL}$ due to the background anisotropy. \\
The amplitude from the ``right" vertex in this case can be written as,
\begin{equation}
A^R({\bf k_1},{\bf k_2},{\bf k_3})=-\frac{\sum_{i<j}k^2_ik^2_j}{\dot{\phi}^2}
\sum_{{\xi_i}=\pm1}(\prod^3_{i=1} G_{\xi_i}(k_i))\frac{1}{\sum_i\xi_ik_i}\left(1-e^{i \eta_0\sum_i\xi_ik_i}\right)  \label{aRf}
\end{equation}
where $G_{\xi_i}(k_i)=F^{p}_{\xi_i}(k_i)$, if $\vec{k}_i$ is in the planar regime, and $G_{\xi_i}(k_i)=F^{np}_{\xi_i}(k_i)$,
if $\vec{k}_i$ is in the non-planar regime. Recall that the functions $F_{\xi_i}(k_i)$ are given as 
\begin{eqnarray}
F_{\xi_i=-1}(k_i)=\frac{H_I^2}{2k^3}(|A^{i}_{+}|^2-A^{i}_{-}{A^{i}_{+}}^{\ast}) \\
F_{\xi_i=1}(k_i)=\frac{H_I^2}{2k^3}(|A^{i}_{-}|^{2}-A^{i}_{+}{A^{i}_{-}}^{\ast})
\end{eqnarray}
For the planar case, the functions $A^i_{\pm}$ are obtained from equation (\ref{apam2}) , while for the non-planar case these are given by (\ref{apam1}). Also, the functions $\tilde{F_{\xi_i}}(k_i)$ are defined as $\tilde{F_{\xi_i}}(k_i)=\frac{2k^3_i}{H^2_I} F_{\xi_i}(k_i)$.\\
The bispectrum has enhancements for both flattened and squeezed triangle configurations, which we catalogue below.\\ \\
\textbf{Flattened Triangle Configurations}\\ \\
\noindent (1) \textit{One non-planar momentum}  \\We choose ${\bf k}_1={\bf k}_2+{\bf k}_3$ for the flattened triangle configuration. There are two different cases to consider:\\ \\
\indent(a) \textit{${\bf k}_1$ and  ${\bf k}_2$ are planar and ${\bf k}_3$ is non-planar}\\ 
In this case $\xi_1=1, \xi_2=-1, \xi_3=-1$. The ``right" amplitude then reduces to 
\begin{equation}
A^R({\bf k},{\bf k},{\bf k})=\frac{3H^6_I}{8\dot{\phi}^2 k^6} \tilde{F}^p_{+}({\bf k}) \tilde{F}^p_{-}({\bf k}) \tilde{F}^{np}_{-}({\bf k})(ik\eta_0)
\end{equation}
Therefore, the change in the $f_{NL}$ parameter is given as,
\begin{eqnarray}
\Delta f^{flat}_{NL} &=& \frac{A^R({\bf k},{\bf k},{\bf k}) + c.c.}{ P^2_{p}({\bf k}) + 2P_{p}({\bf k})P_{np}({\bf k})} \nonumber \\
&\approx &\left(\frac{\dot{\phi}^2_e}{H^2_I}\right)\frac{\sinh{\pi a s k/H_I}\cos{2bk/H_I} }{ (\cosh{\pi a s k/H_I}-\sin{2bk/H_I})(2\sinh{\pi a s k/H_I}  +\cosh{\pi a s k/H_I}-\sin{2bk/H_I})} \nonumber \\
\end{eqnarray}
\indent(b) \textit{${\bf k}_2$ and  ${\bf k}_3$ are planar and ${\bf k}_1$ is non-planar}\\
In this case $\xi_1=-1, \xi_2=1, \xi_3=1$. The ``right" amplitude then reduces to 
\begin{equation}
A^R({\bf k},{\bf k},{\bf k})=\frac{3H^6_I}{8\dot{\phi}^2 k^6} \tilde{F}^{np}_{-}({\bf k}) \tilde{F}^p_{+}({\bf k}) \tilde{F}^{p}_{+}({\bf k})(ik\eta_0)
\end{equation}
Therefore, the change in the $f_{NL}$ parameter is given as,
\begin{eqnarray}
\Delta f^{flat}_{NL} &=& \frac{A^R({\bf k},{\bf k},{\bf k}) + c.c.}{ P^2_{p}({\bf k}) + 2P_{p}({\bf k})P_{np}({\bf k})} \nonumber \\
&\approx &\left(\frac{\dot{\phi}^2_e}{H^2_I}\right)\frac{-\sin{4bk/H_I} + 2e^{-\pi a s k/H_I}\cos{2bk/H_I}}{ (\cosh{\pi a s k/H_I}-\sin{2bk/H_I})(2\sinh{\pi a s k/H_I}  +\cosh{\pi a s k/H_I}-\sin{2bk/H_I})} \nonumber \\
\end{eqnarray}
The total change in $f_{NL}$ is given as the sum of the contributions (a) and (b). In the limit  $s > H_I/k$ we obtain $$\Delta f^{flat}_{NL} \approx \frac{\dot{\phi}^2_e}{H^2_I}e^{-\pi a s k/H_I}$$ so that no appreciable enhancement can be observed for this configuration.\\

\noindent (2) \textit{Two non-planar momenta} \\
We again choose ${\bf k}_1={\bf k}_2+{\bf k}_3$ for the flattened triangle configuration. In this case, ${\bf k}_1$ is planar while ${\bf k}_2,{\bf k}_3$ are non-planar corresponding to $\xi_1=1, \xi_2=-1, \xi_3=-1$.\\
The ``right" amplitude in this case is given by
\begin{equation}
A^R({\bf k},{\bf k},{\bf k})=\frac{3H^6_I}{8\dot{\phi}^2 k^6} \tilde{F}^p_{+}({\bf k}) \tilde{F}^{np}_{-}({\bf k}) \tilde{F}^{np}_{-}({\bf k})(ik\eta_0)
\end{equation}
Therefore, the change in the $f_{NL}$ parameter is given as,
\begin{eqnarray}
\Delta f^{flat}_{NL} &=& \frac{A^R({\bf k},{\bf k},{\bf k}) + c.c.}{ P^2_{p}({\bf k}) + 2P_{p}({\bf k})P_{np}({\bf k})} \nonumber \\
&\approx & \left(\frac{\dot{\phi}^2_e}{H^2_I}\right)\frac{\cos{2bk/H_I} }{(\sinh{\pi a s k/H_I}  +2\cosh{\pi a s k/H_I}-2\sin{2bk/H_I})} \nonumber \\
\end{eqnarray}
For the same reasons as explained in case (1), non-Gaussianity from this configuration is also severely suppressed.\\

\noindent\textbf{Squeezed Triangle Configurations}\\ \\
\noindent (1) \textit{One non-planar momentum} \\ Taking ${\bf k}_1={\bf k}_2+{\bf k}_3$, with $k_3 \ll k_1,k_2$ we again have two cases to consider:\\
\indent(a) \textit{${\bf k}_3$ is non-planar :} This corresponds to the case where $\xi_1=1,\xi_2=-1,\xi_3=-1$, with ${\bf k}_1, {\bf k}_2$ both planar. The ``right" amplitude is given as,
\begin{equation}
A^R({\bf k},{\bf k}_3) \approx \frac{H^6_I}{\dot{\phi}^2 k^2 k^4_3} (1- e^{-i\eta_0 k_3}) \tilde{F}^p_{+}({\bf k}) \tilde{F}^{p}_{-}({\bf k}) \tilde{F}^{np}_{-}({\bf k}_3)
\end{equation}
Therefore, the change in the $f_{NL}$ parameter is given as,
\begin{eqnarray}
\Delta f^{sqzd}_{NL} &=& \frac{A^R({\bf k},{\bf k}_3) + c.c.}{2P_{p}({\bf k})P_{np}({\bf k}_3)} \nonumber \\
&\approx & \left(\frac{\dot{\phi}^2_e}{H^2_I}\right)\frac{k}{k_3} \frac{1}{(\sinh{\pi a s k/H_I} )(\cosh{\pi a s k/H_I}-\sin{2bk/H_I})} [ 2(1-\cos{\eta_0 k_3}) \nonumber \\
&+&e^{\pi a s k/H_I} (-\sin{2bk/H_I}+\sin{(2bk/H_I +\eta_0 k_3 )}) - e^{-\pi a s k/H_I} (\sin{2bk/H_I}+\sin{(2bk/H_I -\eta_0 k_3 )})] \nonumber \\
\end{eqnarray}

\indent(b) \textit{${\bf k}_3$ is planar :} There can be two contributions in this case depending on whether $\xi_3$ is $+1$ or $-1$:\\
\indent(i) \textit{$\xi_3=-1$} \\
In this case,
\begin{equation}
A^R({\bf k},{\bf k}_3) \approx \frac{H^6_I}{\dot{\phi}^2 k^2 k^4_3} (1- e^{-i\eta_0 k_3}) \tilde{F}^p_{+}({\bf k}) \tilde{F}^{np}_{-}({\bf k}) \tilde{F}^{p}_{-}({\bf k}_3)
\end{equation}
The change in the $f_{NL}$ parameter is given as,
\begin{eqnarray}
\Delta f^{sqzd}_{NL} &=& \frac{A^R({\bf k},{\bf k}_3) + c.c.}{P_{p}({\bf k})P_{p}({\bf k}_3)+ P_{np}({\bf k})P_{p}({\bf k}_3)} \nonumber \\
&\approx & \left(\frac{\dot{\phi}^2_e}{H^2_I}\right)\frac{k}{k_3} \frac{(1- e^{-i\eta_0 k_3})(e^{\pi a s_3 k_3/H_I}+ie^{i2bk_3/H_I})(e^{-\pi a s k/H_I}- ie^{-i2bk/H_I}) + c.c.}{(\cosh{\pi a s_3 k_3/H_I}- \sin{2bk_3/H_I})(\sinh{\pi a s k/H_I}+\cosh{\pi a s k/H_I}-\sin{2bk/H_I})} \nonumber \\
\end{eqnarray}

(ii)\textit{$\xi_3=1$} \\
In this case,
\begin{equation}
A^R({\bf k},{\bf k}_3) \approx \frac{H^6_I}{\dot{\phi}^2 k^2 k^4_3} (1- e^{-i\eta_0 k_3}) \tilde{F}^{np}_{-}({\bf k}) \tilde{F}^{p}_{+}({\bf k}) \tilde{F}^{p}_{+}({\bf k}_3)
\end{equation}
The change in the $f_{NL}$ parameter is given as,
\begin{eqnarray}
\Delta f^{sqzd}_{NL} &=& \frac{A^R({\bf k},{\bf k}_3) + c.c.}{P_{p}({\bf k})P_{p}({\bf k}_3)+ P_{np}({\bf k})P_{p}({\bf k}_3)} \nonumber \\
&\approx & \left(\frac{\dot{\phi}^2_e}{H^2_I}\right)\frac{k}{k_3} \frac{(1- e^{-i\eta_0 k_3})(e^{-\pi a s_3 k_3/H_I}-ie^{-i2bk_3/H_I})(e^{-\pi a s k/H_I}- ie^{-i2bk/H_I}) + c.c.}{(\cosh{\pi a s_3 k_3/H_I}- \sin{2bk_3/H_I})(\sinh{\pi a s k/H_I}+\cosh{\pi a s k/H_I}-\sin{2bk/H_I})} \nonumber \\
\end{eqnarray}
The net change in the $f_{NL}$ parameter in the case of one non-planar momentum with ${\bf k}_3$ being planar is the sum of the contributions (i) and (ii).\\ 

\noindent (2) \textit{Two non-planar momenta} There is only one case of interest in this category, which corresponds to $\xi_1=1, \xi_2=-1, \xi_3=-1$ - ${\bf k}_2, {\bf k}_3$ are non-planar.\\
\begin{equation}
A^R({\bf k},{\bf k}_3) \approx \frac{H^6_I}{\dot{\phi}^2 k^2 k^4_3} (1- e^{-i\eta_0 k_3}) \tilde{F}^{p}_{+}({\bf k}) \tilde{F}^{np}_{-}({\bf k}) \tilde{F}^{np}_{-}({\bf k}_3)
\end{equation}
The change in the $f_{NL}$ parameter is given as,
\begin{eqnarray}
\Delta f^{sqzd}_{NL} &=& \frac{A^R({\bf k},{\bf k}_3) + c.c.}{P_{np}({\bf k})P_{np}({\bf k}_3)+ P_{p}({\bf k})P_{np}({\bf k}_3)} \nonumber \\
&\approx & \left(\frac{\dot{\phi}^2_e}{H^2_I}\right)\frac{k}{k_3} \frac{e^{-\pi a s k/H_I}(1-\cos{\eta_0 k_3})- \sin{2bk/H_I} +\sin{(2bk/H_I +\eta_0 k_3)}}{(\sinh{\pi a s k/H_I}+\cosh{\pi a s k/H_I}-\sin{2bk/H_I})} \nonumber \\
\end{eqnarray}
In the regime $sk/H_I > 1$, one therefore has the following leading order contribution to the $f_{NL}$ parameter for all the above "mixed" cases:
\begin{equation}
\Delta f^{sqzd}_{NL} \approx \left(\frac{\dot{\phi}^2_e}{H^2_I}\right)\frac{k}{k_3} e^{-\pi a s k/H_I}
\end{equation}
which is precisely the same as the contribution of the squeezed triangle configuration of purely planar modes (\ref{planarsqpure}). Choosing $\frac{\dot{\phi}^2_e}{H^2_I} \sim 10^{-2}$, $\frac{k}{k_3}|_{max} \sim 200$
and $s k/H_I \sim 2$, we have $\Delta f^{sqzd}_{NL}  \sim O(10^{-2})$. These two categories of squeezed triangle configurations maximize the enhancement in the bispectrum due to early time anisotropy.\\ 


\section{Conclusion} This note presents an analytical approach for studying the signature of a strongly anisotropic pre-inflationary phase in the spectrum of late time cosmological perturbations for a range of high momentum modes which typically exit the horizon after isotropization. Starting with the previously known case of 
non-planar modes ($k_x \sim k_y \sim k_z$) \cite{Dey:2011mj}, we have worked out in complete detail the case of the planar modes ($k_x \ll k_y,k_z$). While non-planar modes always admit a WKB description at early times,  WKB condition for planar modes  is violated at very early times  and restored later in course of their evolution. In either case,  the late time dynamics of the modes is characterized by an excited state (built on the standard Bunch-Davis vacuum by a Bogoliubov transformation) which carries all the information about the pre-inflationary anisotropic phase. The problem of space-time anisotropy at early times can, therefore, be recast into a problem of quasi de Sitter evolution with a non-standard vacuum state at late times. We therefore study the correlation functions (two and three point functions) of curvature perturbation in this excited state and calculate the power spectrum and the leading order contributions to the $f_{NL}$ parameter for both planar and non-planar modes due to anisotropy. In addition, we compute the leading order contributions to the $f_{NL}$ parameter in cases where the 3-point correlation function involves  both planar and non-planar modes.\\

In the power spectrum for non-planar modes, the contribution of anisotropy is essentially perturbative in nature, since the corresponding (leading order) correction term in the spectral index for these modes is suppressed by a power of $e^{-t_e/t_{iso}}$. As a result, for $t_e/t_{iso} \sim O(10)$, for example, the correction due to anisotropy is severely suppressed. However, in the case of planar modes, the power spectrum for a generic wavenumber $k$ and parameter $s$ (defined as $s =k_x/k$), strongly breaks scale invariance. In the limit $s \ll H_I/k$,  the power spectrum appears to diverge as $P(k) \sim f(k)/s$, which agrees with predictions from numerical computations in \cite{Gumrukcuoglu:2007bx}. Also the range of $s$ (for  a given $k$) for which the aforementioned $s$- dependence of the power spectrum is valid shrinks with growing $k$. However, from considerations of validity of perturbation theory, we show that the inequality $sk/H_I > 1$ must be satisfied and therefore the regime $s \ll H_I/k$ is not physically relevant. \\

For generic modes in the allowed regime $sk/H_I >1$, the analytical formula for the spectral index suggests that the effect of the initial anisotropy might be substantial. We compute the resulting statistical correlation between different multipoles of the CMB temperature anisotropies and numerically calculate the signal-to-noise ratio for individual multipoles for a physically reasonable range of values of $H_I$ as an estimate of how large  this effect can be. This analysis shows that the signal-to-noise ratio is pretty low even for the maximum allowed value for $H_I$. \\

The bispectrum for non-planar modes computed in the non-standard vacuum state is enhanced in the flattened triangle configuration ($k_2 \approx k_3 \approx k_1/2$) as well as the squeezed triangle configuration ($k_3 \ll k_1 \approx k_2$), with the latter yielding a larger $f_{NL}$ parameter. However, in both cases, the $f_{NL}$ parameter is suppressed by factors of $e^{-3t_e/2t_{iso}}$ in addition to the slow-roll parameter, which constrain the non-Gaussianity in the scalar perturbation spectrum to be extremely small. For planar modes, the bispectrum is similarly enhanced for the flattened triangle as well as the squeezed triangle configuration. In each case we derive the $f_{NL}$ parameter for a generic planar mode of wavenumber $k$ and parameter $s$ and study its behavior in the regime $sk/H_I > 1$ . The squeezed configuration again gives a larger enhancement in the spectrum compared to the flat configuration. However, 
$\Delta f^{sqzd}_{NL} \approx  (\frac{\dot{\phi}^2_e}{H^2_I})(\frac{k}{k_3}) e^{-\pi a s k/H_I}$ involves strong exponential suppression as a result of which it can at best be of $O(10^{-2})$.\\
In the regime of scale-invariant power spectrum i.e. $sk/H_I \gg 1$, $\Delta f^{sqzd}_{NL}$ is vanishingly small as before, because of larger exponential suppression.The condition of scale invariance, therefore, naturally leads to unobservable non-Gaussianity for the planar modes.\\

Finally, we also analyze non-Gaussianity originating from three-point correlation functions involving both planar and non-planar modes. As before, one expects enhancements from the flattened as well as the squeezed triangle configuration with the squeezed triangle configurations producing larger non-Gaussianity. The leading order contribution of the squeezed triangle configurations in the "mixed" case to the $f_{NL}$ parameter turns out to be the same as the purely planar case:  $\Delta f^{sqzd}_{NL} \approx (\frac{\dot{\phi}^2_e}{H^2_I})\frac{k}{k_3} e^{-\pi a s k/H_I}$. For the choice of parameters described in the previous section $\Delta f^{sqzd}_{NL}|_{max} \sim O(10^{-2})$.\\

To summarize, our computation suggest that primordial anisotropy will not produce an observable $f_{NL}$ in the near future. 

\acknowledgments
EDK thanks Assaf Ben-David for tweaked CMBEASY outputs and useful discussions.

\newpage

\section{Appendix A: Non-planar Modes}
In this section, we study the evolution of the non-planar modes for a single, massless scalar field (referred to as $f$ in the main body of the paper) minimally coupled to gravity in an axially symmetric anisotropic space-time, with a positive cosmological constant. In particular, we construct a WKB solution for these modes at early times and show how one can match it with the general solution at late times to obtain equations (\ref{apam1}). The treatment essentially follows \cite{Kim:2010wra}.\\
We consider the following action for the scalar field,
\begin{equation}
S= - \int d^4x \sqrt{-g}\left(\frac{1}{2}g^{\mu\nu}\partial_{\mu}\phi\partial_{\nu}\phi+ V\right), \hspace{4ex} (M_p^2\equiv 1)
\end{equation}
where the background metric is chosen to be an axially symmetric version of the
Bianchi I metric:
\begin{equation}
ds^2=-dt^2 + \exp{(2\rho)}(dx)^2 + \exp{(2\beta)}(dy^2+dz^2)
\end{equation}
where  $\rho,\beta$ are known functions of time:
\begin{eqnarray}
\rho& =& \frac{1}{3}\ln{\tanh^{2}{\frac{3H_I t}{2}}\sinh{3H_I t}} \nonumber \\
\beta& =& \frac{1}{3}\ln{\frac{\sinh{3H_I t}}{\tanh{\frac{3H_I t}{2}}}} \label{classeom}
\end{eqnarray}
with $H_I=\sqrt{\frac{V}{3}}$.\\
Define $\rho=\alpha - 2\beta_+, \beta=\alpha+\beta_+$ and a new ``time'' coordinate $\tau$, analogous to the conformal time in the isotropic limit,  as,
\begin{equation}
d\tau =\frac{dt}{e^{3 \alpha}}
\end{equation}
From equation $(\ref{classeom})$ and the definition of the coordinate $\tau$, one can derive, $e^{3\alpha}= e^{\rho+2\beta}=\sinh{3H_I t}= \frac{1}{\sinh{-3H_I \tau}}$.\\
It can be easily seen that as $t$ varies from $0^+$ to $\infty$, $\tau$ varies from $-\infty$ to $0^-$. In this time coordinate, the equation of motion for a mode $\phi_k$ is given by,
\begin{equation}
\left(\frac{d^2}{d\tau^2}+\omega(\tau)^2\right)\phi_k=0 \label{modeeom}
\end{equation}
The frequency squared is given as,
\begin{equation}
\omega(\tau)^2=\frac{2^{\frac{4}{3}}k^2}{x^{\frac{4}{3}}}(1-r^2 x) \label{omegasq}
\end{equation}
where $r^2=\frac{k_y^2+k_z^2}{k^2}$ and $x(\tau)=1-e^{6H_{I}\tau}=\exp{(-6\alpha)}(\sqrt{\exp{(6\alpha)}+1}-1)$. 
Evidently, $x(\tau)$ varies from 1 to 0 as $\tau$ changes from $-\infty$ to $0$.\\
Equation ($\ref{modeeom}$) has a WKB solution:
\begin{equation}
\phi_{WKB}=\frac{1}{\sqrt{2\tilde{\omega}}}\exp{[-i\int^{\tau}_{\tau_0}d\tau^{'}
\tilde{\omega}]} \label{WKB}
\end{equation}
where $\tilde{\omega}$ has to be determined from the equation,
\begin{equation}
\tilde{\omega}^2=\omega^2 -\frac{1}{2}\left(\frac{\tilde{\omega}_{,\tau\tau}}{\tilde{\omega}}-\frac{3\tilde{\omega}^2_{,\tau}}{2\tilde{\omega}^2}\right)
\end{equation}
The WKB approximation holds as long as the WKB parameter
\begin{eqnarray}
\varepsilon &= & \left|\frac{\frac{d\omega^2}{d\tau}}{\omega^3}\right| \nonumber \\ \nonumber \\
& = & \frac{H_{I}}{k}\frac{1-x(\tau)}{(x/2)^{1/3}(1-r^2x(\tau))^{1/2}}\left(\frac{3}{1-r^2x(\tau)}+1\right)\ll 1
\end{eqnarray} 
The choice of the WKB solution above is obviously equivalent to imposing a particular initial condition on the 
modes of the scalar field at early times. This can be seen directly by analyzing the early time behavior of the classical solution for the scalar field. Firstly note that $\tilde{\omega} \approx \omega \to k_x$ in the limit $\tau\to -\infty$ (or $t \to 0+$) and in this limit $\tau$ and $t$ are related as $\tau=\frac{1}{3H_I} \ln{\frac{3H_I t}{2}}$. Therefore, in the early time limit, the time-dependence of the WKB solution is given as follows,
\begin{equation}
\phi_{WKB} \approx \frac{1}{\sqrt{2 k_x}} \exp[-i\frac{k_x}{3H_I} \log{\frac{3H_I t}{2}}] \sim t^{-i\frac{k_x}{3H_I} }
\end{equation}
Now consider the equation of motion of the scalar field at early times,
\begin{equation}
\ddot{\phi_k} +\frac{1}{t}\dot{\phi_k} + (k_y^2 +k_z^2 + \frac{k_x^2}{3Vt^2})\phi_k=0 \label{earlytime}
\end{equation}
Define $z= \ln{\sqrt{k_y^2 +k_z^2 } t}$, so that the equation reduces to 
\begin{equation}
\phi^{''}_k(z) + (e^{2z} +\frac{k_x^2}{3V} )\phi_k(z)=0
\end{equation}
In the limit $t \to 0+$, $z \to -\infty$, so that the exponential term drops out of the above equation and we obtain a solution of the form, 
\begin{equation}
\phi_k(z)= A(k) e^{-i\frac{k_x}{\sqrt{3V}}z } + B(k) e^{i\frac{k_x}{\sqrt{3V}}z }
\end{equation}
Choosing $B(k)=0$, we find that $\phi_k(z) \sim t^{-i\frac{k_x}{3H_I} }$, confirming that the WKB solution has the same time-dependence at early times as expected from the classical solution subject to a certain initial condition.\\
In fact, equation (\ref{earlytime}) has the general solution, 
\begin{equation}
\phi_k(t)= C_1(k) H^{(1)}_{ik_x/\sqrt{3V}} (\sqrt{k_y^2 +k_z^2 } t) + C_2(k) H^{(2)}_{ik_x/\sqrt{3V}} (\sqrt{k_y^2 +k_z^2 } t)  
\end{equation}
Using $H^{(1)}_{i\nu} (z)=\frac{1}{\sinh{\pi \nu}} (J_{i\nu}(z) e^{\pi \nu} - J_{-i\nu}(z))$ and $H^{(2)}_{i\nu} (z)=\frac{1}{\sinh{\pi \nu}} (-J_{i\nu}(z) e^{-\pi \nu}+ J_{-i\nu}(z))$, we can rewrite the general solution as,
\begin{equation}
\phi_k(t)=A(k) J_{ik_x/\sqrt{3V}}(\sqrt{k_y^2 +k_z^2 } t) + B(k) J_{-ik_x/\sqrt{3V}}(\sqrt{k_y^2 +k_z^2 } t)
\end{equation}
Therefore, the WKB solution corresponds to imposing the initial condition $A(k)=0$ and choosing $B(k)$ appropriately, as $J_{-ik_x/\sqrt{3V}}(\sqrt{k_y^2 +k_z^2 } t)$ has the same time-dependence at early times as our WKB solution.\\

We will be interested in the large momentum regime of non-planar wavenumbers, i.e. $k_i\gg H$, implying that the factor $(1-r^2x(\tau))$ in the denominator doesn't vanish anywhere (since both $r$ and $x$ are fractions). In the regime $x\approx 1$ (i.e. early times), the WKB condition obviously holds for any momentum (unless $r \sim 1$ which is the planar case discussed in Appendix B). In \cite{Kim:2010wra}, it was shown that the condition holds  for high-momentum modes as long as  $\frac{k}{H} \gg \exp{\alpha(t)}$. Therefore, the time at which the WKB solution should be matched with the late-time de Sitter solution has a natural choice, $\tau_{*}$, such that,
\begin{equation}
e^{\alpha(\tau_{*})}=\sqrt{\frac{k}{H_I}}
\end{equation}
The above equation implicitly states that $\tau_{*}$ corresponds to late times when $e^{\alpha(\tau_*)} \gg 1$. In terms of real time, this condition implies that $e^{\alpha(t_*)} \approx e^{H_I t_*} \gg 1$, or $t_* > t_{iso} \sim \frac{1}{H_I}$. Also, if $t_e$ denotes the time of horizon exit for a given mode of wavenumber $k$, we have $e^{H_I t_e} =\frac{k}{H}$. This suggests a simple relation between $t_*$ and $t_e$, viz.
\begin{equation}
t_e \approx 2 t_*
\end{equation}

Returning to the problem of matching the modes - since $k\gg H_I$, $x(\tau_*) \approx \exp{(-6\alpha(\tau_*))} \approx 0$. Therefore the WKB solution around $\tau=\tau_*$ is given by expanding equation ($\ref{WKB}$) around $x=0$ and then plugging in the values of $x$ and the frequencies at $t=t_*$.
The solution can be expanded in powers of $\varepsilon=\sqrt{\frac{H_I}{k}}$ and one needs to retain terms to the order at which the direction dependence first appears. It turns out that it is sufficient to retain terms up to the order $\varepsilon^3$ and to this order $x(\tau_*)$ and the frequencies are given as,
\begin{equation}
x(t_*)=2(H_I/k)^{3/2}(1-(H_I/k)^{3/2})
\end{equation}
\begin{equation}
\omega_*=\frac{k^2}{H_I} \, \left[1+(\frac{2}{3}-r^2)(\frac{H_I}{k})^{3/2}\right]
\end{equation}
\begin{equation}
\tilde{\omega}_*^2=\omega_*^2 \left(1-2\frac{H_I}{k}\right)
\end{equation}

Now, in the de Sitter regime the solution to equation (7.5) is given as,
\begin{equation}
\phi_k=A_+\phi_+(\tau) + A_-\phi_-(\tau)
\end{equation}
where the modes $\phi_{\pm}$ are given as,
\begin{equation}
\phi_{\pm}(\tau)=\left(\pm i + \frac{ik}{H_I}(-3H_I\tau)^{1/3}\right)\exp{\left(\pm \frac{ik}{H_I}(-3H_I\tau)^{1/3}\right)}
\end{equation}
Matching the de Sitter solution with the WKB solution at $\tau=\tau_*$, we obtain the following equations for $A_+$ and $A_{-}$,
\begin{eqnarray}
A_+\phi'_+(\tau_*) + A_-\phi'_-(\tau_*) & = & -\left[\left(1-\frac{H_I}{k}\right)-i\sqrt{\frac{H_I}{k}}\left(1-\frac{H_I}{2k}\right)\right]\sqrt{\frac{\omega(t_*)}{2}}\\
A_+\phi_+(\tau_*) + A_-\phi_-(\tau_*)& = -i& \left(1+\frac{H_I}{2k}\right)\sqrt{\frac{1}{2\omega_*}}
\end{eqnarray}
where we have absorbed an overall phase in the definition of $A_+$
and $A_-$. Solving for $A_{\pm}$ from the above equations, we have,
\begin{eqnarray}
A_{+} & = & \frac{\varepsilon^3}{2\sqrt{2H_I}}\left[\left(2-\varepsilon^2 \right)+2i\varepsilon\left(\frac{\varepsilon^2}{2}-1\right) +O(\varepsilon^4)\right]\exp{\left(\frac{-i}{\varepsilon}\right)}\\
A_{-}& = & -\frac{\varepsilon^3}{2\sqrt{2H_I}}\left[\left(\frac{2}{3}-r^2 \right)\varepsilon^3 +O(\varepsilon^4) \right]\exp{\left(\frac{i}{\varepsilon}\right)}
\end{eqnarray}
One can easily verify that these coefficients obey the normalization condition,
\begin{equation}
|A_+|^2-|A_-|^2=\frac{H_I^2}{2k^3}
\end{equation}
In the main body of the paper, we have redefined $A_{\pm} \to \sqrt{\frac{H_I^2}{2k^3}} A_{\pm}$, such that the normalization condition reads,
\begin{equation}
|A_+|^2-|A_-|^2=1
\end{equation}

\section{Appendix B : Planar Modes}
In this section, we discuss the solution of the classical equation of motion for the planar modes of a scalar field evolving in the axially symmetric Bianchi I background specified in Appendix A. The computation is analogous to the non-planar case and hence we only provide a summary of the main results. For details, the reader may refer to \cite{Kim:2011pt}.\\
As before, we start from
\begin{equation}
\left(\frac{d^2}{d\tau^2}+\omega(\tau)^2\right)\phi_k=0
\end{equation}
where the frequency squared is given as $\omega(\tau)^2=\frac{2^{\frac{4}{3}}k^2}{x^{\frac{4}{3}}}(1-r^2 x)$ with $r^2=\frac{k_y^2+k_z^2}{k^2}$ and $x(\tau)=1-e^{6H_{I}\tau}=\exp{(-6\alpha)}(\sqrt{\exp{(6\alpha)}+1}-1)$ and $x(\tau)$ varies from 1 to 0 as $\tau$ changes from $-\infty$ to $0$.\\
Now, recall that the WKB parameter is given as,
\begin{equation}
\varepsilon =  \frac{H_{I}}{k}\frac{1-x(\tau)}{(x/2)^{1/3}(1-r^2x(\tau))^{1/2}}\left(\frac{3}{1-r^2x(\tau)}+1\right)
\end{equation} 

For $k_x \ll k_y, k_z$, the denominator $(1-r^2x(\tau))^{1/2}$ can be an extremely small fraction at early times (when $x(\tau) \to 1$) and the WKB parameter exceeds unity. Therefore, in this regime one needs to solve the equation above directly and impose some initial condition on the modes. This is subsequently matched at later times with the WKB solution, which, in turn, is matched to the late time de Sitter solution.\\
At early times, the frequency squared can be decomposed as $\omega^2 \approx 2^{\frac{4}{3}}(k_x^2+ \delta \omega^2e^{6H_I\tau}) $, where $\delta \omega^2 \approx k_y^2 +k_z^2$.
 the equation of motion in the limit $\tau \to -\infty $ therefore has the general solution,
\begin{equation}
\phi^{1}_k= C_1(k) J_{-iak_x/H_I}(qe^{3H_I \tau}) + C_2(k) J_{iak_x/H_I}(qe^{3H_I \tau}) 
\end{equation}
with $q=a \sqrt{\delta \omega^2}/H_I=a \sqrt{k_y^2 +k_z^2}/H_I$ ( $a=\frac{2^{2/3}}{3}$ is a numerical constant).\\
Following \cite{Kim:2011pt}, we fix the initial condition as,
\begin{equation}
\phi^{1}_k= \sqrt{\frac{\pi}{6H_I \sinh{\pi a k_x/H_I}}} J_{-iak_x/H_I}(qe^{3H_I \tau}) \label{earlysol}
\end{equation}
In the limit $\tau \to -\infty$, this solution assumes the form of an incoming waveform. By noting that $\Gamma(1+i\nu)=\sqrt{\frac{\pi \nu}{\sinh{\pi \nu}}}$, we have
\begin{eqnarray}
\phi^{1}_k(\tau \to -\infty)&=& \sqrt{\frac{\pi}{6H_I \sinh{\pi a k_x/H_I}}}\frac{1}{\Gamma(1-ia k_x/H_I)}(q e^{3 H_I \tau}/2)^{-ia k_x/H_I} \\ \nonumber
&=&\frac{1}{\sqrt{6 a k_x}} \exp{[-3iak_x\tau +i\psi]}
\end{eqnarray}
where $\psi$ is a time-independent phase factor defined as $e^{i\psi}=(a\sqrt{k_y^2 + k_z^2}/H_I)^{-ia k_x/H_I}$.\\

At later times, the WKB condition becomes valid again and the solution to the equation of motion is given as,
\begin{equation}
\phi^{2}_k=\frac{B_{+}}{\sqrt{2\tilde{\omega}}}\exp{[-i\int^{\tau}_{\tau_0}d\tau^{'}\tilde{\omega}]} +\frac{B_{-}}{\sqrt{2\tilde{\omega}}}\exp{[i\int^{\tau}_{\tau_0}d\tau^{'}\tilde{\omega}]}
\end{equation}
where $\tilde{\omega}$ has to be determined from the equation $\tilde{\omega}^2=\omega^2 -\frac{1}{2}\left(\frac{\tilde{\omega}_{,\tau\tau}}{\tilde{\omega}}-\frac{3\tilde{\omega}^2_{,\tau}}{2\tilde{\omega}^2}\right)$, as before.\\
Let $\tau_0$ be the time at which $\phi^{1}_k$ and $\phi^{2}_k$ are matched to determine $B_{\pm}$ and this has to be chosen such that both the early time solution (equation $\ref{earlysol}$) and the WKB solution are valid. A suitable choice is given by,
\begin{equation}
e^{3H_I \tau_{0}} \simeq \frac{1}{a}\sqrt{\frac{H_I}{(\delta\omega^2)^{\frac{1}{2}}}} \approx \frac{1}{a}\sqrt{\frac{H_I}{k}}
\end{equation}
The matching procedure leads to the following solution for $B_{\pm}$,
\begin{eqnarray}
B_{+} \approx \frac{1}{\sqrt{(1-e^{-2\pi a k_x/H_I})}} e^{i(\pi/4 -\sqrt{\frac{(\delta \omega^2)^{1/2}}{H_I}} )} \\
B_{-} \approx  \frac{e^{-\pi a k_x/H_I}}{\sqrt{(1-e^{-2\pi a k_x/H_I})}} e^{-i(\pi/4 -\sqrt{\frac{(\delta \omega^2)^{1/2}}{H_I}} )}
\end{eqnarray}
Finally, we match the WKB solution to the de Sitter solution
\begin{eqnarray}
\phi^{(3)}_{k}(\tau)&=&\frac{H_I}{2k^{3/2}} A_{+}\left(i + \frac{ik}{H_I}(-3H_I\tau)^{1/3}\right)\exp{\left( \frac{ik}{H_I}(-3H_I\tau)^{1/3}\right)} \nonumber \\
&+& \frac{H_I}{2k^{3/2}} A_{-}\left(-i + \frac{ik}{H_I}(-3H_I\tau)^{1/3}\right)\exp{\left(- \frac{ik}{H_I}(-3H_I\tau)^{1/3}\right)}
\end{eqnarray}
at a matching time $\tau=\tau_*$, such that $e^{\alpha(\tau_*)}=\sqrt{\frac{k}{H_I}}$ and this leads to the following solution for $A_+$ and $A_-$,
\begin{eqnarray}
A_{+} & =& \frac{e^{i(\pi/4 - b k/H_I)}}{\sqrt{1-e^{-2\pi (a s k/H_I)}}} \nonumber\\ \nonumber\\
A_{-} & = &  \frac{e^{-\pi (a s k/H_I)}e^{-i(\pi/4 - b k/H_I)}}{\sqrt{1-e^{-2\pi (a s k/H_I)}}}
\end{eqnarray}
where $s= k_x/k \ll 1$ is a measure of "planarity" of the mode while $a=\frac{2^{2/3}}{3}$ and $b=\frac{2^{2/3}\sqrt{\pi} \Gamma(1/3)}{3\Gamma(5/6)}$ are numerical constants.\\

The above expressions for $A_{\pm}$ completely specify the late-time vacuum state of the \\
theory. 2-point and the 3-point correlation functions of the scalar field in the ``in-in" formalism can now be computed with respect to this vacuum.\\

\end{document}